\documentclass[12pt]{article}

\usepackage{graphicx}
\usepackage{amssymb}
\usepackage{amsmath}
\usepackage{amsfonts}
\usepackage{latexsym}
\usepackage{color}
\usepackage{verbatim}
\usepackage{float}

\newcommand{\dslash}{\hskip.1ex\hbox to 0pt{/\hss}\kern-.1ex\partial}

\begin{document}
\begin{titlepage}

\begin{flushright}
\end{flushright}

\begin{center}
{\large \bf Screening of an electrically charged particle in a two-dimensional two-component plasma at $\Gamma=2$}
\end{center}

\vspace{1.5mm}

\begin{center}
{Alejandro Ferrero\footnote{{\tt a.ferrero@uniandes.edu.co}} and Gabriel T\'ellez\footnote{{\tt gtellez@uniandes.edu.co}}}

\vspace{1.5mm}
{\sl Departamento de F\'isica} \\
{\sl Universidad de los Andes} \\
{\sl Bogot\'a, Colombia} \\

\end{center}

\vspace{0.2ex}

\medskip

\centerline{\bf Abstract}

We consider the thermodynamic effects of an electrically charged
impurity immersed in a two-dimensional two-component plasma, composed
by particles with charges $\pm e$, at temperature $T$, at coupling
$\Gamma=e^2/(k_B T)=2$, confined in a large disk of radius
$R$. Particularly, we focus on the analysis of the charge density, the
correlation functions, and the grand potential. Our analytical results
show how the charges are redistributed in the circular geometry
considered here. When we consider a positively charged impurity, the
negative ions accumulate close to the impurity leaving an excess of
positive charge that accumulates at the boundary of the disk. Due to
the symmetry under charge exchange, the opposite effect takes place
when we place a negative impurity. Both the cases in which the
impurity charge is an integer multiple of the particle charges in the
plasma, $\pm e$, and a fraction of them are considered; both
situations require a slightly different mathematical treatments,
showing the effect of the quantization of plasma charges. The bulk and
tension effects in the plasma described by the grand potential are not
modified by the introduction of the charged particle. Besides the
effects due to the collapse coming from the attraction between
oppositely charged ions, an additional topological term appears in the
grand potential, proportional to $-n^2\ln(mR)$, with $n$ the
dimensionless charge of the particle. This term modifies the central
charge of the system, from $c=1$ to $c=1-6n^2$, when considered in the
context of conformal field theories.
\end{titlepage}

\section{Introduction}
A two-dimensional two-component plasma (TCP) is a gas composed by two
kinds of electrically charged particles (with charges $\pm e$)
interacting through an electrostatic potential. In two dimensions, the
Coulomb interaction between two unit charges separated by a distance
$r$ is $-\ln (r/L)$ with $L$ an irrelevant length. The value of the
charges, together with the inverse temperature factor
$\beta=(k_BT)^{-1}$ determine a coupling constant given by
$\Gamma=\beta e^2$.

The value of $\Gamma$ is important to determine the behavior and
stability of the system. For low temperatures ($\Gamma\ge2$), the
thermal energy is not enough to avoid the collapse of oppositely
charged ions and the point particles must be replaced by hard disks of
radius $a$. The point particle view, however, can still be kept when
$\Gamma<2$ because the thermal fluctuations allow to avoid such
collapse. For the value $\Gamma=2$, which we will assume in this
analysis, the classical Coulomb gas is mathematically isomorph to a
quantum free Fermi field at zero temperature
\cite{ref-corn}. Moreover, analytical solutions for the grand
potential, densities and correlation functions can be found for this
particular value of the coupling constant.

The analogy between the two-component plasma and the sine-Gordon
quantum field theory has been exploited to obtain results for the
thermodynamic properties of the Coulomb gas under some conditions when
$\Gamma<2$ \cite{ST,SJ-tcp-metal,Samaj-tcp-diele}. When $\Gamma>2$, an
infinite-order transition that take place when $\Gamma=4$ at low
density (the Kosterlitz-Thouless transition) has been studied
\cite{Koster}; additionally, exact results for the plasma for values
of $\Gamma$ larger than 2, but close enough to 2, have also been found
\cite{kalinay,tellez3}.

In the particular case $\Gamma=2$ several studies have been done in
the past: the one-component plasma (a system composed
by only one kind of particles immersed in an oppositely charged
background) with adsorbing impurities \cite{cuatro,Cornu-ocp-ad}, and
the two component plasma with adsorbing boundaries
\cite{siete,Merchan-Tellez-jabon-anillos} and adsorbing impurities
\cite{ferrero}.

In this work, we study the introduction of an electrically charged
impurity into the plasma. The plasma occupies a disk region of radius
$R$. The impurity is an impenetrable disk of radius $r_0$ and charge
$q=n e$ located at the center of the disk domain that contains the
plasma. The electrostatic potential generated by the impurity, at a
distance $r$ from it, can be written as $V(r)=-ne\ln(r/L)$, where $n$
is the dimensionless charge of the impurity. From this point on, we
will understand that the charge $q$ of the impurity is ``integer'' if
$q$ is an integer multiple of $e$, i.e., $q=ne$, with $n\in\mathbb
Z$. On the other hand, a ``non-integer'' charge means that
$n\notin\mathbb Z$ and so there is a fractional part in the charge of
the impurity that cannot be compensated by the other particles in the
gas. Using the electrostatic potential just described, we can find the
fugacities for positive and negative particles; they are
\begin{eqnarray}
m_s(\mathbf r)=
m(\mathbf r)e^{-s \beta e V(r)}=
m(\mathbf r)\left(\frac{r}{L}\right)^{2sn},
\end{eqnarray}
where $s=\pm1$, and $m(\mathbf{r})=0$ for $|\mathbf{r}|=r<r_0$ and
$m(\mathbf{r})=m$ for $r>r_0$.

This model can be applied to some physical systems. The
present concentrical circular geometry can be seen as a transversal
cut of a cylindrical geometry. The location of the impurity at the
center of the system generates the redistribution of charge that
induces the accumulation of positive or negative ions around the
impurity and the boundary of the plasma.
Some proteins and other components in biological systems could be
modeled as large cylinders with hard core effects and electrical
charges spread along their length. To be able to describe the
effective interactions between these charged entities immersed in an
electrolyte is an important task towards understanding its physical
properties.

Although the screening of charges in an electrolyte is fairly well
understood in the mean field regime, $\Gamma\to 0$, described by the
Poisson--Boltzmann equation~\cite{debye1923theory, fuoss1951potential,
  manning1969limiting1}, the intermediate and strong coupling regimes
remain elusive. There has been several efforts towards the
understanding of the strong coupling regime~\cite{RB96,
  naji2004attraction, naji2005electrostatic, trizac1, SaTr11PRE,
  Mallarino13}. However, most of the strong-coupling techniques, such
as the Wigner strong coupling approach \cite{trizac1, SaTr11PRE}, are
only targeted at the one-component plasma, and cannot directly be
applied to a multicomponent plasma such as the one studied
here. Thus, the present work tries to contribute to the understanding
of the screening beyond the mean field by studying the intermediate
coupling $\Gamma=2$.

There has been previous studies of charged impurities in the
two-dimensional two-component plasma beyond mean field, for
$0<\Gamma<2$, but limited to point impurities in the bulk. In these
works, the use of the form factors of the sine-Gordon field theory
allows to obtain the behavior of the density profiles at large
distances from the impurity~\cite{Samaj05} and using the product
operator expansion, one can obtain the short distance behavior of the
densities~\cite{Tellez05}. However, the complete form of the density
profiles for the whole range of distances is unknown. At $\Gamma=2$,
the equivalent field theory to the two-component plasma is a free
fermion one, therefore it will be possible to obtain more explicit
results in the present case, and also consider the effets of the
nonzero size of the impurity ($r_0\neq 0$), and the confinement of the
plasma ($R\ < \infty$).

This paper is organized as follows: in section \ref{form} we briefly discuss the theoretical background for a two-dimensional two-component plasma at $\Gamma=2$. Section \ref{green} discusses the Green's functions that generate the density and correlations of the plasma. The density of the gas and the total charge are discussed in sections \ref{dens} and \ref{char} respectively. Chapter \ref{poten} explains the calculation of the grand potential. We finally summarize our conclusions and discuss the results in section \ref{conc}.

\section{Formalism}\label{form}
In 1989, Cornu and Jancovici developed a general formalism to study a
two-dimensional two-component gas at $\Gamma=e^2\beta=2$
\cite{ref-corn}. Within this model, the two-dimensional position of a
particle $\mathbf{r}=(x,y)$ can be written in complex coordinates as
$z=x+iy$. As explained before, when $\Gamma=2$ the point particles
must be replaced by hard spheres of radius $a$; its inverse $a^{-1}$
could also be viewed as an ultraviolet cutoff in momentum space. In
the limit $a\rightarrow0$, the logarithm of the grand partition
function $\Xi$ for this system takes the form~\cite{ref-corn}
\begin{eqnarray}
\ln\Xi=\textrm{Tr}\left[\ln\left(\not\!\partial+m_+(\mathbf r)\frac{1+\sigma_z}{2}+m_-(\mathbf r)\frac{1-\sigma_z}{2}\right)-\ln\not\!\partial\right].
\end{eqnarray}
In a two-dimensional system,
$\dslash=\sigma_x\partial_x+\sigma_y\partial_y$, where $\sigma_x$,
$\sigma_y$ and $\sigma_z$ are the Pauli matrices. The terms
$m_{\pm}(\mathbf r)=m(\mathbf r)e^{-\beta U_{\pm}(\mathbf r)}$ are
fugacities that take into account the effects of an electric external
potential that generates electrostatic energies given by $U_{\pm}(r)$
for positive and negative particles. The term $m(\mathbf r)$ includes
hard-core effects and other non-electrical contributions.  The
calculation of the grand potential, $\Omega=- k_B T\ln\Xi$, is
performed by finding the eigenvalues $\{\lambda_k\}$ of the system
\cite{ref-corn}
\begin{eqnarray}\label{eq:pot1}
&&\left[\not\!\partial-\frac{1}{\lambda}\left(
\begin{array}{cc}
m_+(\mathbf r) & 0\\ 0 & m_-(\mathbf r)
\end{array}\right)\right]\psi(\mathbf r)=0, \,\,\,\textrm{with}\,\,\,\,
\psi(\mathbf r)=\left(\begin{array}{c}g(\mathbf r) \\ f(\mathbf r)\end{array}\right)
\end{eqnarray}
a two-component spinor. Then the grand potential is given by
\begin{eqnarray}\label{eq:pot}
\beta\Omega=-\sum_k\ln(1+\lambda_k).
\end{eqnarray}
These two last equations show the equivalence between this system and
a free Fermi gas \cite{ref-corn}.  The pressure $p$, in the
thermodynamic limit, is given by
\begin{eqnarray}
\beta p=\frac{\partial\ln \Xi}{\partial A}=-\frac{1}{A}\beta\Omega,
\end{eqnarray}
where $A$ is the area of the gas.
The truncated $p$-body densities are given by:
\begin{eqnarray}\label{eq:correl}
\rho_{s_1,s_2\dots s_p}^{(p)T}(\mathbf r_1,\mathbf r_2,\dots,\mathbf r_p)&=&(-1)^{p+1}m_{s_1}(\mathbf r_1)m_{s_2}(\mathbf r_2)\dots m_{s_p}(\mathbf r_p)\nonumber\\
&\times& \sum_{i_1,i_2\dots,i_p}G_{s_{i_1},s_{i_2}}(\mathbf r_{i_1},\mathbf r_{i_2})\dots
G_{s_{i_p},s_{i_1}}(\mathbf r_{i_p},\mathbf r_{i_1}),
\end{eqnarray}
where the summations run over cycles $\{i_1,i_2\dots,i_n\}$.
$G_{s_1s_2}(\mathbf r_1,\mathbf r_2)$ are matrix elements that satisfy the set of differential equations
\begin{eqnarray}\label{eq:greenf}
\left(\begin{array}{cc}m_+(\mathbf r_1) & 2\partial_{z_1}\\2\partial_{\bar z_1} & m_-(\mathbf r_1)\end{array}\right)
\left(\begin{array}{cc}G_{++}(\mathbf r_1,\mathbf r_2) & G_{+-}(\mathbf r_1,\mathbf r_2)\\
G_{-+}(\mathbf r_1,\mathbf r_2) & G_{--}(\mathbf r_1,\mathbf r_2)\end{array}\right)=\delta(\mathbf r_1-\mathbf r_2).
\end{eqnarray}
For the particular cases $n=1$ and $n=2$ we have
\begin{eqnarray}\label{eq-13}
\rho_{s}^{(1)}(\mathbf r)\!\!\!&=&\!\!\!m_s(\mathbf r)G_{ss}(\mathbf r,\mathbf r),\nonumber\\
\rho_{s_1s_2}^{(2)T}(\mathbf r_1,\mathbf r_2)\!\!\!&=&\!\!\!
-m_{s_1}(\mathbf r_1)m_{s_2}(\mathbf r_2)G_{s_1s_2}(\mathbf r_1,\mathbf r_2)G_{s_2s_1}(\mathbf r_2,\mathbf r_1).
\end{eqnarray}
Since the gas is not stable under the collapse of particles of opposite signs, the results will depend on the cutoff distance given by $a$. While the limit $a\rightarrow0$ can be taken for some quantities such as the correlation functions, the charge density and other quantities will depend on such distance.

\section{The Green functions}\label{green}

We will now study the effects of a charged particle of radius $r_0$
located at the center of the plasma with a potential given by
$U(r)=-ne^2\ln(r/L)$. This particle is thus immersed in a large
electrical system confined in a large disk of radius $R$.  Defining
$g_{s_1s_2}=e^{-is_1 U(r_1)}G_{s_1s_2}e^{-is_2 U(r_2)}$ and using
Eq.~(\ref{eq:greenf}), we obtain the set of differential equations
(for $r_0<r<R$)
\begin{eqnarray}\label{eq:diff-green}
\Bigg\{m^2-\nabla^2_{1}
-\frac{2in}{r_1^2}\frac{\partial}{\partial\theta_1}
+\frac{n^2}{r_1^2}\Bigg\}
g_{\pm\pm}(\mathbf r_1,\mathbf r_2)=m\delta(\mathbf r_1-\mathbf r_2),\\
g_{-+}(\mathbf r_1,\mathbf r_2)=-m^{-1}\left[2\partial_{\bar z_1}-n\frac{e^{i\theta_1}}{r_1}\right]g_{++}(\mathbf r_1,\mathbf r_2),\\
g_{+-}(\mathbf r_1,\mathbf r_2)=-m^{-1}\left[2\partial_{z_1}+n\frac{e^{-i\theta_1}}{r_1}\right]g_{--}(\mathbf r_1,\mathbf r_2).
\end{eqnarray}
We now propose a particular solution in terms of a Fourier expansion
with $l$-modes $g_{ss'}^l(\mathbf r_1,\mathbf
r_2)=e^{il\theta_1}g_{ss'}^l(\tilde r_1, \tilde{\mathbf{r}}_2)$, where we defined
the dimensionless variables $\tilde r_{1,2}=mr_{1,2}$. In terms of
these variables and using the stated Fourier decomposition, we see
that the modes $g_{ss'}^l(\tilde r_1,\tilde{\mathbf{r}}_2)$ obey the differential equations
\begin{eqnarray}\label{eq:diff-rad}
  \Bigg\{\tilde r_1^2\frac{d}{d\tilde r_1^2}+\tilde r_1\frac{d}{d\tilde r_1}-\Big[(l+n)^2+\tilde r_1^2\Big]\Bigg\}
  g_{\pm\pm}^l(\tilde r_1,\tilde{\mathbf{r}}_2)=m\delta(\tilde r_1-\tilde r_2)e^{-il\theta_2},\\
    g_{-+}^l(\tilde r_1,\tilde{\mathbf{r}}_2)=-e^{i\theta_1}\left[\frac{d}{d\tilde r_1}
      -\frac{1}{\tilde r_1}(l+n)\right]g_{++}(\tilde r_1,\tilde{\mathbf{r}}_2),\label{eq:diff-rada}\\
    g_{+-}^l(\tilde r_1,\tilde{\mathbf{r}}_2)=-e^{-i\theta_1}\left[\frac{d}{d\tilde r_1}
      +\frac{1}{\tilde r_1}(l+n)\right]g_{--}(\tilde r_1,\tilde{\mathbf{r}}_2).\label{eq:diff-radb}
\end{eqnarray}
The Fourier modes can thus be written in the form $g_{\pm\pm}^l(\tilde
r_1,\tilde{\mathbf{r}}_2)=A_l(\tilde{\mathbf{r}}_2)K_{l+n}(\tilde
r_1)+B_l(\tilde{\mathbf{r}}_2)I_{l+n}(\tilde r_1)$, where $I_l(x)$ and
$K_l(x)$ are the modified Bessel functions of first and second kind
and $A_l(\tilde{\mathbf{r}}_2)$ and $B_l(\tilde{\mathbf{r}}_2)$ are
coefficients to be determined by the boundary conditions. We thus have
(for $r_0<r_1<R$ and $r_1<r_2$)
\begin{eqnarray}\label{eq:general}
g_{\pm\pm}\!\!\!\!&=&\!\!\!\!
\sum_l e^{il\theta_1}\left[A_l^{(\pm\pm)}K_{l+n}(\tilde r_1)
+B_l^{(\pm\pm)}I_{l+n}(\tilde r_1)\right],\\
g_{\mp\pm}\!\!\!\!&=&\!\!\!\!
\sum_l e^{i(l\pm1)\theta_1}\left[A_{l}^{(\pm\pm)}K_{l+n\pm1}(\tilde r_1)
-B_{l}^{(\pm\pm)}I_{l+n\pm1}(\tilde r_1)\right],
\end{eqnarray}
and something similar for $r_2<r_1$ where the coefficients for that region will be called $A'^{(\pm\pm)}_l$ and $B'^{(\pm\pm)}_l$.

In the regions $r<r_0$ and $r>R$ we have that the fugacity vanishes so $m=0$. Therefore, the differential equations for $G_{s_1s_2}$ imply that
\begin{eqnarray}\label{eq:ges}
&&G_{++}^{\,l}=u_l^{(++)}e^{il\theta_1}r_1^{l},\,\,\,\,\,\,\,\,G_{+-}^{\,l}=u_l^{(+-)}e^{i(l-1)\theta_1}r_1^{l-1},\\\label{eq:ges1}
&&G_{--}^{\,l}=u_l^{(--)}e^{il\theta_1}r_1^{-l},\,\,\,\,\,\,G_{-+}^{\,l}=u_l^{(-+)}e^{i(l+1)\theta_1}r_1^{-(l+1)},
\end{eqnarray}
where the coefficients $u^{(s_1s_2)}_l$ have to be determined for each one
of the two regions. Eqs.~(\ref{eq:diff-rad})$-($\ref{eq:diff-radb}),
(\ref{eq:ges}) and (\ref{eq:ges1}) imply the following boundary
conditions:
\begin{enumerate}
\item $g_{\pm\pm}$ must be continuous at $r_1=r_2$.
\item $g_{\pm\mp}$ must be discontinuous at $r_1=r_2$.
\item $G_{s_1s_2}$ must be finite at $r=0$.
\item $G_{s_1s_2}$ must vanish at $r=\infty$.
\end{enumerate}
While the first and the second conditions are given by the delta
distribution in Eq. (\ref{eq:diff-rad}) and provide relations for the
coefficients $A_l$ and $B_l$ in the regions $r_0<r_{1,2}<R$, the third
and fourth conditions provide the extra conditions by stating which
$u_l^{(s_1s_2)}$ coefficients must vanish.

From now on we will use the notations $F(\tilde r_j)\equiv F^{(j)}$ and $F(\tilde R)\equiv F^{(R)}$. The continuity conditions at $r_{1}=r_0$ imply
\begin{eqnarray}\label{cond1}
A_l^{(--)}K_{l+n}^{(0)}+B_{l}^{(--)}I_{l+n}^{(0)}\!\!\!&=&\!\!\!0,\,\,\,l\ge1,\\
A_l^{(--)}K_{l+n-1}^{(0)}-B_{l}^{(--)}I_{l+n-1}^{(0)}\!\!\!&=&\!\!\!0,\,\,\,l\le0,\\
A_l^{(++)}K_{l+n+1}^{(0)}-B_{l}^{(++)}I_{l+n+1}^{(0)}\!\!\!&=&\!\!\!0,\,\,\,l\ge0,\\
A_l^{(++)}K_{l+n}^{(0)}+B_{l}^{(++)}I_{l+n}^{(0)}\!\!\!&=&\!\!\!0,\,\,\,l\le-1.
\end{eqnarray}
Additionally, the continuity conditions at $r_{1}=R$ demand:
\begin{eqnarray}\label{cond2}
A_l'^{(--)}K_{l+n}^{(R)}+B_l'^{(--)}I_{l+n}^{(R)}\!\!\!&=&\!\!\!0,\,\,\,l\le0,
\\
A_l'^{(--)}K_{l+n-1}^{(R)}-B_l'^{(--)}I_{l+n-1}^{(R)}\!\!\!&=&\!\!\!0,\,\,\,l-1\ge0,
\\
A_l'^{(++)}K_{l+n+1}^{(R)}-B_l'^{(++)}I_{l+n+1}^{(R)}\!\!\!&=&\!\!\!0,\,\,\,l+1\le0,
\\
A_l'^{(++)}K_{l+n}^{(R)}+B_l'^{(++)}I_{l+n}^{(R)}\!\!\!&=&\!\!\!0,\,\,\,l\ge 0.
\end{eqnarray}
On the other hand, the two conditions at $r_1=r_2$ are
\begin{eqnarray}\label{cond3}
A_l^{(\pm\pm)}K_{l+n}^{(2)}+B_l^{(\pm\pm)}I_{l+n}^{(2)}
-A_l'^{(\pm\pm)}K_{l+n}^{(2)}-B_l'^{(\pm\pm)}I_{l+n}^{(2)}\!\!\!&=&\!\!\!0,\\
A_l'^{(\pm\pm)}K_{l+n\pm1}^{(2)}-A_l^{(\pm\pm)}K_{l+n\pm1}^{(2)}
-B_l'^{(\pm\pm)}I_{l+n\pm1}^{(2)}+B_l^{(\pm\pm)}I_{l+n\pm1}^{(2)}\!\!\!&=&\!\!\!
\frac{m}{2\pi\tilde r_2}e^{-il\theta_2}.\label{cond3a}
\end{eqnarray}
The particular Green functions are thus found by solving for the coefficients given in Eqs.~(\ref{cond1})$-$(\ref{cond3a}).
Using the definitions 
\begin{equation}
  t_{l}^{(x)}\equiv\frac{K_{l}(x)}{I_l(x)}
\,,
\end{equation}
and $\theta_{12}\equiv\theta_1-\theta_2$, and
writing $n$ as $n=k+\nu$, where $k$ and $\nu$ are the integer and fractional parts of $n$ respectively, we find
\begin{eqnarray}\label{ges1}
g_{++}\!\!\!&=&\!\!\!
\frac{m}{2\pi}e^{-ik\theta_{12}}\Big[\sum_le^{il\theta_{12}}I_{l+\nu}^{(<)}K_{l+\nu}^{(>)}
+\Delta_{++}^{(\nu,\tilde r_0,\tilde R)}\Big],
\\
g_{--}\!\!\!&=&\!\!\!
\frac{m}{2\pi}e^{-ik\theta_{12}}\Big[
\sum_le^{il\theta_{12}}I_{l+\nu}^{(<)}K_{l+\nu}^{(>)}
+\Delta_{--}^{(\nu,\tilde r_0,\tilde R)}\Big],
\end{eqnarray}
where the notation $(<)$ and $(>)$ means that we choose between the smaller and larger between $r_1$ and $r_2$ respectively and
\begin{eqnarray}\label{delta1}
\Delta_{++}^{(\nu,\tilde r_0,\tilde R)}\!\!\!&=&\!\!\!
-\sum_{l=-\infty}^{k-1}e^{il\theta_{12}}\frac{[t_{l+\nu}^{(0)}]^{-1}K_{l+\nu}^{(1)}K_{l+\nu}^{(2)}}
{1+[t_{l+\nu}^{(0)}]^{-1}t_{l+\nu+1}^{(R)}}
+\sum_{l=k}^{\infty}e^{il\theta_{12}}
\frac{[t_{l+\nu+1}^{(0)}]^{-1}K_{l+\nu}^{(1)}K_{l+\nu}^{(2)}}
{1+[t_{l+\nu+1}^{(0)}]^{-1}t_{l+\nu}^{(R)}}\nonumber\\
\!\!\!&&\!\!\!
-\sum_{l=-\infty}^{k-1}e^{il\theta_{12}}
t_{l+\nu+1}^{(R)}\frac{[t_{l+\nu}^{(0)}]^{-1}(I_{l+\nu}^{(1)}K_{l+\nu}^{(2)}+K_{l+\nu}^{(1)}I_{l+\nu}^{(2)})-I_{l+\nu}^{(1)}I_{l+\nu}^{(2)}}
{1+[t_{l+\nu}^{(0)}]^{-1}t_{l+\nu+1}^{(R)}}
\nonumber\\
\!\!\!&&\!\!\!
-\sum_{l=k}^{\infty}e^{il\theta_{12}}
t_{l+\nu}^{(R)}\frac{[t_{l+\nu+1}^{(0)}]^{-1}(I_{l+\nu}^{(1)}K_{l+\nu}^{(2)}+K_{l+\nu}^{(1)}I_{l+\nu}^{(2)})+I_{l+\nu}^{(1)}I_{l+\nu}^{(2)}}
{1+[t_{l+\nu+1}^{(0)}]^{-1}t_{l+\nu}^{(R)}},
\\\label{delta2}
\Delta_{--}^{(\nu,\tilde r_0,\tilde R)}\!\!\!&=&\!\!\!
\sum_{l=-\infty}^{k}e^{il\theta_{12}}\frac{[t_{l+\nu-1}^{(0)}]^{-1}K_{l+\nu}^{(1)}K_{l+\nu}^{(2)}}
{1+[t_{l+\nu-1}^{(0)}]^{-1}t_{l+\nu}^{(R)}}
-\sum_{l=k+1}^{\infty}e^{il\theta_{12}}
\frac{[t_{l+\nu}^{(0)}]^{-1}K_{l+\nu}^{(1)}K_{l+\nu}^{(2)}}
{1+[t_{l+\nu}^{(0)}]^{-1}t_{l+\nu-1}^{(R)}}\nonumber\\
\!\!\!&&\!\!\!
-\sum_{l=-\infty}^{k}e^{il\theta_{12}}
t_{l+\nu}^{(R)}\frac{[t_{l+\nu-1}^{(0)}]^{-1}(I_{l+\nu}^{(1)}K_{l+\nu}^{(2)}+K_{l+\nu}^{(1)}I_{l+\nu}^{(2)})+I_{l+\nu}^{(1)}I_{l+\nu}^{(2)}}
{1+[t_{l+\nu-1}^{(0)}]^{-1}t_{l+\nu}^{(R)}}
\nonumber\\
\!\!\!&&\!\!\!-\sum_{l=k+1}^{\infty}e^{il\theta_{12}}
t_{l+\nu-1}^{(R)}\frac{[t_{l+\nu}^{(0)}]^{-1}(I_{l+\nu}^{(1)}K_{l+\nu}^{(2)}+K_{l+\nu}^{(1)}I_{l+\nu}^{(2)})-I_{l+\nu}^{(1)}I_{l+\nu}^{(2)}}
{1+[t_{l+\nu}^{(0)}]^{-1}t_{l+\nu-1}^{(R)}}.
\end{eqnarray}
The functions $g_{\pm\mp}$ are found using Eqs.~(\ref{eq:diff-rada})
and (\ref{eq:diff-radb}).

From this point we should emphasize the difference between the cases $n\in\mathbb Z$ ($\nu=0$) and $n\notin\mathbb Z$ ($\nu\neq0$). Mathematically,  when $l\in\mathbb Z$, $I_{l}(x)=I_{-l}(x)$ but $I_{-(l+\nu)}(x)=I_{l+\nu}(x)+d_{\,\nu}^{\,l}K_{l+\nu}(x)$, with $d_{\,\nu}^{\,l}\equiv\frac{2}{\pi}(-1)^{l}\sin(\pi\nu)$.

Up to a phase factor, the term
$\frac{m}{2\pi}\sum_le^{il\theta_{12}}I_{l+\nu}^{(<)}K_{l+\nu}^{(>)}$
in last equations reproduces the contribution for the unperturbed
plasma $g^0_{ss'}(\mathbf r_1,\mathbf r_2)=\frac{m}{2\pi}K_0(|\tilde
r_1-\tilde r_2|)$ in the case $\nu=0$.  For $\nu\neq0$ we separate
this term by writing
\begin{eqnarray}\label{g0}
\frac{m}{2\pi}\sum_le^{il\theta_{12}}I_{l+\nu}^{(<)}K_{l+\nu}^{(>)}=g^0_{ss'}(\mathbf r_1,\mathbf r_2)
+\frac{m}{2\pi}\sum_le^{il\theta_{12}}(I_{l+\nu}^{(<)}K_{l+\nu}^{(>)}-I_{l}^{(<)}K_{l}^{(>)}).
\end{eqnarray}

The remaining contributions, defined as $\Delta_{s_1s_2}^{(\nu,\tilde
  r_0,\tilde R)}$, depend on $\nu$, $r_0$ and $R$ and describe effects
purely generated by the impurity.

\section{Density Profile}\label{dens}
The density of the positive and negative particles in the plasma can
easily be evaluated using the Green's functions found in the previous
section. Using Eq.~(\ref{eq:correl}), we can verify that
\begin{eqnarray}\label{enes}
\rho_{\pm}(\mathbf r)=mg_{\pm\pm}(\mathbf r,\mathbf r).
\end{eqnarray}
We can also define the charge density (divided by $e$), $\rho_{t}(\mathbf
r)$, as the difference between the density of positive and negative
particles; it is given by
\begin{eqnarray}\label{enet}
\rho_{t}(\mathbf r)=m[\,g_{++}(\mathbf r,\mathbf r)-g_{--}(\mathbf r,\mathbf r)\,].
\end{eqnarray}
Since the case $\nu=0$ is a particular situation of the more general
case $\nu\neq0$, we will assume the most general situation and use the
limit $\nu\rightarrow0$ in order to find the solutions for the
particular case in which the impurity has an integer charge. Remember
that in the unperturbed case $\rho^0_{\pm}(\mathbf r)\equiv
\rho^0\displaystyle\mathop{=}_{a\to 0}\frac{m^2}{2\pi}\left(\,\ln\frac{2}{ma}-\gamma\right)+O(1)$
\cite{ref-corn}, where $a$ is the cutoff and $\gamma$ is the
Euler-Mascheroni constant.

\subsection{General Case}
The density profile can be found using
Eqs.~(\ref{delta1})$-$(\ref{enet}). For computational reasons it is
better to eliminate the negative modes in Eqs. (\ref{delta1}) and
(\ref{delta2}) in favor of only positive ones. Using
$I_{-(l+\nu)}(x)=I_{l+\nu}(x)+d_{\nu}^{\,l}K_{l+\nu}(x)$ we notice
that
\begin{eqnarray}
[t_{-l+\nu}^{(0)}]^{-1}=[t_{l-\nu}^{(0)}]^{-1}-d_{\nu}^{\,l},\hspace{1cm}
t_{-l+\nu+1}^{(R)}=\frac{t_{l-\nu-1}^{(R)}}{1+d_{\nu}^{\,l}t_{l-\nu-1}^{(R)}}.
\end{eqnarray}
We use a definition for $\nu$ in which $-1<\nu<1$. For example, if
$n=7/2$ we have that $k=3$ and $\nu=1/2$, but for $n=-7/2$ we have
$k=-3$ and $\nu=-1/2$. We thus use the notation $\nu_{\pm}$ to
indicate the cases when $n\ge0$ and $n<0$ respectively. Then we find
\begin{eqnarray}\label{densidades}
\frac{2\pi}{m^2}\rho_{+}^{(\nu_+,\tilde r_0,\tilde R)}(\mathbf r)\!\!\!&=&\!\!\!
\frac{2\pi}{m^2}\tilde \rho^0_\nu
-\sum_{l=0}^{k-1}\frac{[t_{l+\nu}^{(0)}]^{-1}[\,K_{l+\nu}^{(r)}]^2}
{1+[t_{l+\nu}^{(0)}]^{-1}t_{l+\nu+1}^{(R)}}
-\sum_{l=0}^{k-1}
t_{l+\nu+1}^{(R)}\frac{2[t_{l+\nu}^{(0)}]^{-1}I_{l+\nu}^{(r)}K_{l+\nu}^{(r)}-[\,I_{l+\nu}^{(r)}]^{2}}
{1+[t_{l+\nu}^{(0)}]^{-1}t_{l+\nu+1}^{(R)}}
\nonumber\\
\!\!\!&&\!\!\!
-\sum_{l=1}^{\infty}\frac{[t_{l-\nu}^{(0)}]^{-1}[\,K_{l-\nu}^{(r)}]^2}
{1+[t_{l-\nu}^{(0)}]^{-1}t_{l-\nu-1}^{(R)}}
-\sum_{l=1}^{\infty}
t_{l-\nu-1}^{(R)}\frac{2[t_{l-\nu}^{(0)}]^{-1}I_{l-\nu}^{(r)}K_{l-\nu}^{(r)}-[\,I_{l-\nu}^{(r)}]^{2}}
{1+[t_{l-\nu}^{(0)}]^{-1}t_{l-\nu-1}^{(R)}}
\nonumber\\
\!\!\!&&\!\!\!
+\sum_{l=k}^{\infty}\frac{[t_{l+\nu+1}^{(0)}]^{-1}[\,K_{l+\nu}^{(r)}]^{2}}
{1+[t_{l+\nu+1}^{(0)}]^{-1}t_{l+\nu}^{(R)}}
-\sum_{l=k}^{\infty}
t_{l+\nu}^{(R)}\frac{2[t_{l+\nu+1}^{(0)}]^{-1}I_{l+\nu}^{(r)}K_{l+\nu}^{(r)}+[\,I_{l+\nu}^{(r)}]^{2}}
{1+[t_{l+\nu+1}^{(0)}]^{-1}t_{l+\nu}^{(R)}},
\\\label{densidades1}
\frac{2\pi}{m^2}\rho_{-}^{(\nu_+,\tilde r_0,\tilde R)}(\mathbf r)\!\!\!&=&\!\!\!
\frac{2\pi}{m^2}\tilde \rho^0_\nu
+\sum_{l=0}^{k}\frac{[t_{l+\nu-1}^{(0)}]^{-1}[\,K_{l+\nu}^{(r)}]^{2}}
{1+[t_{l+\nu-1}^{(0)}]^{-1}t_{l+\nu}^{(R)}}
-\sum_{l=0}^{k}t_{l+\nu}^{(R)}\frac{2[t_{l+\nu-1}^{(0)}]^{-1}I_{l+\nu}^{(r)}K_{l+\nu}^{(r)}+[\,I_{l+\nu}^{(r)}]^{2}}
{1+[t_{l+\nu-1}^{(0)}]^{-1}t_{l+\nu}^{(R)}}
\nonumber\\
\!\!\!&&\!\!\!
+\sum_{l=1}^{\infty}\frac{[t_{l-\nu+1}^{(0)}]^{-1}[\,K_{l-\nu}^{(r)}]^{2}}
{1+[t_{l-\nu+1}^{(0)}]^{-1}t_{l-\nu}^{(R)}}
-\sum_{l=1}^{\infty}
t_{l-\nu}^{(R)}\frac{2[t_{l-\nu+1}^{(0)}]^{-1}I_{l-\nu}^{(r)}K_{l-\nu}^{(r)}+[\,I_{l-\nu}^{(r)}]^{2}}
{1+[t_{l-\nu+1}^{(0)}]^{-1}t_{l-\nu}^{(R)}}
\nonumber\\
\!\!\!&&\!\!\!
-\sum_{l=k+1}^{\infty}
\frac{[t_{l+\nu}^{(0)}]^{-1}[\,K_{l+\nu}^{(r)}]^{2}}
{1+[t_{l+\nu}^{(0)}]^{-1}t_{l+\nu-1}^{(R)}}
-\sum_{l=k+1}^{\infty}t_{l+\nu-1}^{(R)}\frac{2[t_{l+\nu}^{(0)}]^{-1}I_{l+\nu}^{(r)}K_{l+\nu}^{(r)}
-[\,I_{l+\nu}^{(r)}]^{2}}{1+[t_{l+\nu}^{(0)}]^{-1}t_{l+\nu-1}^{(R)}},
\end{eqnarray}
with
\begin{eqnarray}
\tilde \rho^0_\nu=\frac{m^2}{2\pi}\sum_{l=0}^{\infty}I_{l+\nu}^{(r)}K_{l+\nu}^{(r)}
+\frac{m^2}{2\pi}\sum_{l=1}^{\infty}I_{l-\nu}^{(r)}K_{l-\nu}^{(r)}.
\end{eqnarray}
When $\nu<0$ we can obtain the solutions using the symmetry relations $\rho_+(\nu_+,k)=\rho_-(-\nu_-,-k)$ and $\rho_+(-\nu_-,-k)=\rho_-(\nu_+,k)$. Notice that
$\tilde \rho^0_0=\rho^0$ when $\nu=0$.

\begin{figure}
\centering
\includegraphics[scale=0.85]{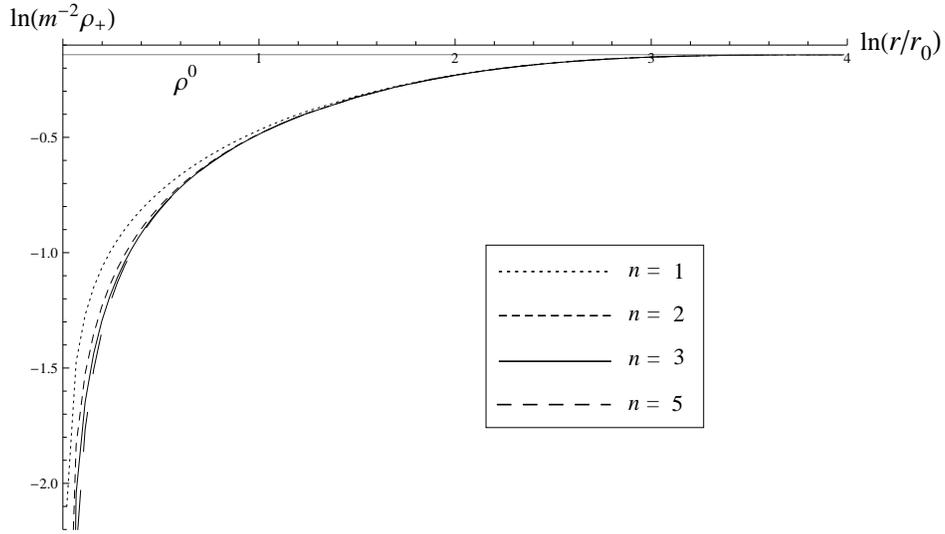}
\caption{Density of the positive charge for four different
  integer impurities located at the origin. The size of the impurity
  is $mr_0=0.05$ and the size of the disk is $mR=30$. The density
  becomes zero as $r\rightarrow r_0$. We used $m^{-2}\rho^0=0.87.$}
\label{fig1}
\end{figure}
\begin{figure}
\centering
\includegraphics[scale=0.85]{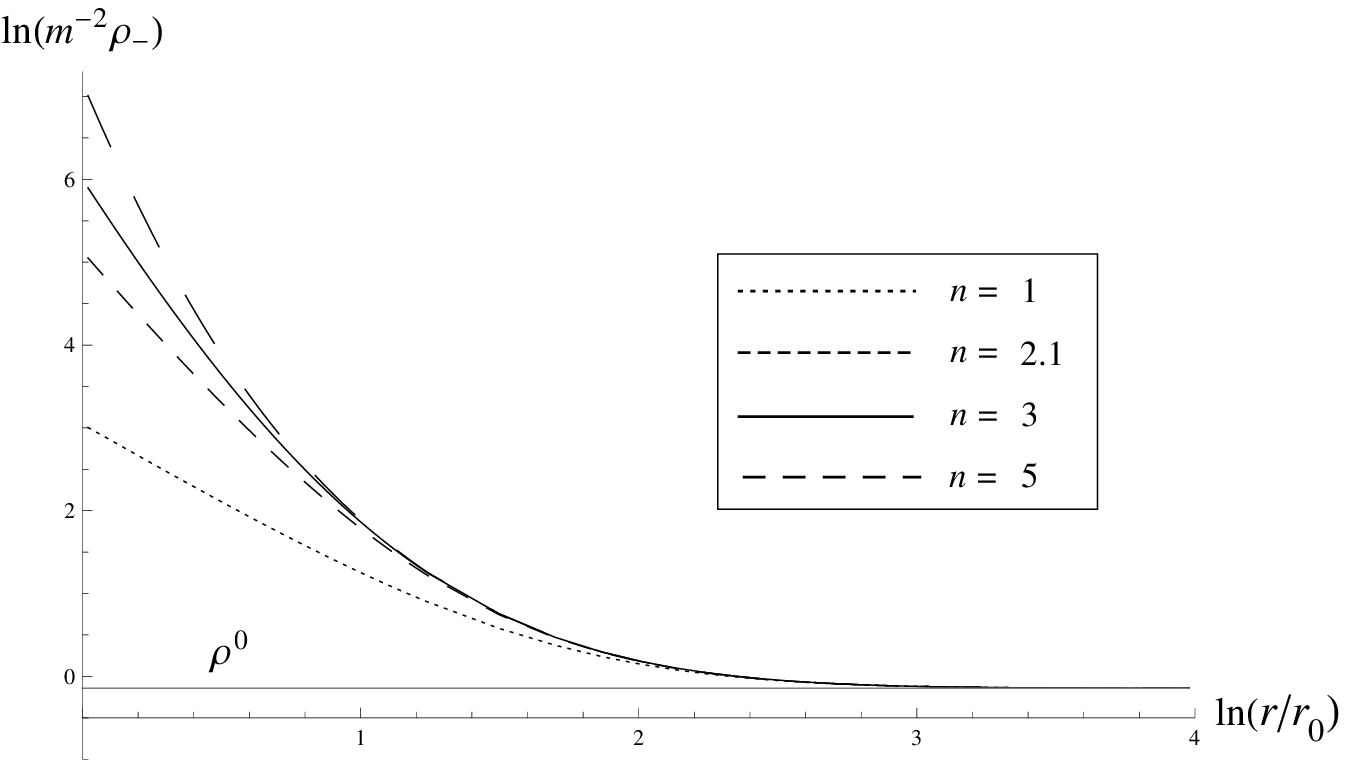}
\caption{Density of the negative charge for four different
  impurities located at the origin. The size of the impurity is
  $mr_0=0.05$ and the size of the disk is $mR=30$. We used $m^{-2}\rho^0=0.87.$}
\label{fig2}
\end{figure}
\begin{figure}
\centering
\includegraphics[scale=0.85]{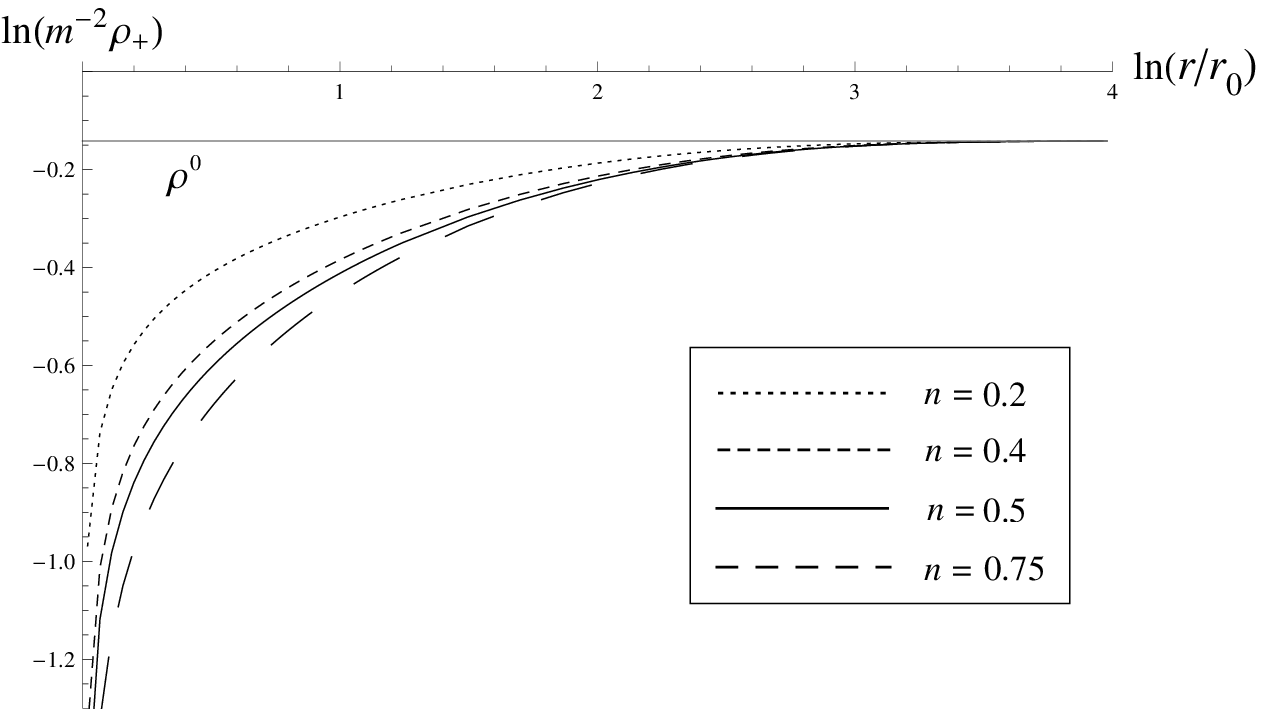}
\caption{Density of the positive charge for four different fractional impurities located at the origin. The size of the impurity is $mr_0=0.05$ and the size of the disk is $mR=30$. We used $m^{-2}\rho^0=0.87.$}
\label{fig3}
\end{figure}
\begin{figure}[t]
\centering
\includegraphics[scale=0.85]{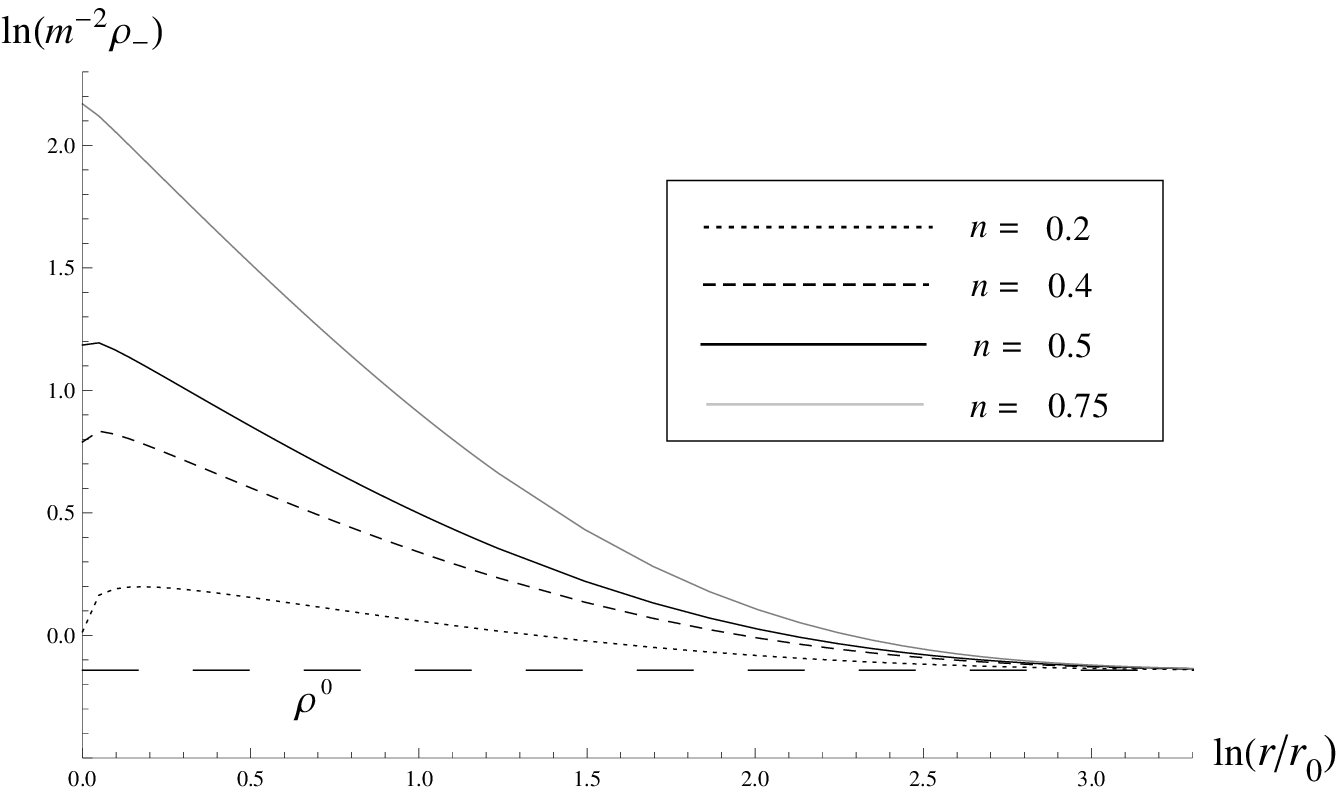}
\caption{Density of the negative charge for four different fractional
  impurities located at the origin. The size of the impurity is
  $mr_0=0.05$ and the size of the disk is $mR=30$. We used $m^{-2}\rho^0=0.87.$}
\label{fig4}
\end{figure}

\begin{figure}[t]
\centering
\includegraphics[scale=0.85]{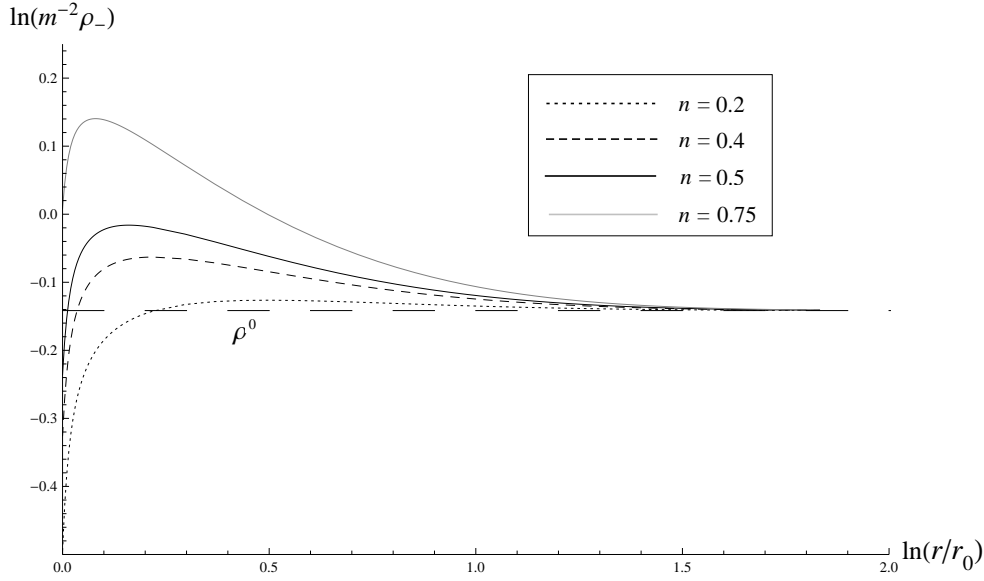}
\caption{Density of the negative charge for four different fractional
  impurities located at the origin. The size of the impurity is
  $mr_0=0.4$ and the size of the disk is $mR=30$. Notice that the
  negative density close to the impurity decreases due to the
  reduction of counter-ions when $r_0$ increases. A large impurity
  avoids the strong collapse of counter-ions due to its hard core
  effects; this effect does not occur as $r_0\to0$.  We used
  $m^{-2}\rho^0=0.87.$}
\label{fig4a}
\end{figure}

Figures~\ref{fig1}, \ref{fig2}, \ref{fig3} and \ref{fig4} show the
density of positive and negative particles for different values of
charge impurity $n>0$. The density of positive particles vanishes near
$r_0$ due to the repulsive interaction with the impurity carrying a
positive charge. On the contrary the density of negative particles
increases in an effort to screen the impurity. Far from the impurity
both densities converge to the bulk density $\rho^{0}$. For $n<1$, the
behavior of the negative density in the presence of a fractional
impurity presents an interesting effect which takes place near $r_0$
approaching it from above (see figure \ref{fig4}). The amount of
negative ions slightly decreases after taking a maximum value in the
neighborhood of the impurity. This behavior is more notable for
smaller values of $\nu$. This effect, however, seems to be a property
satisfied only by charges $n<1$ and not by any non-integer impurity;
as can be seen in figure \ref{fig2}, such behavior is absent for
$n=2.1$. The electrostatic interaction between the impurity and a
negative charge of the plasma can be caracterized by a coupling
constant given by $\Gamma_n=n\beta e^2$ (see table \ref{table1} for
more details). If the impurity radius is very small $r_0\to0$, then
when $n\ge 1$ ($\Gamma_n\ge 2$) there will be a collapse of the
negative particles against the charged impurity, therefore a very
large value of $\rho_{-}(r)$ as $r\to0$. On the other hand, if $n<1$
the system is stable against that collapse and so the value of the
negative density at the origin decreases.  (Such behavior is described
in table \ref{table2}.)  A fingerprint of these two regimes can be
seen in figures \ref{fig2}, \ref{fig4} and \ref{fig4a} which show a
change of behavior in the density when the impurity charge changes
from $n<1$ to $n>1$. A similar observation was made in the analysis
when $r_0=0$ and $\Gamma<2$~\cite{Tellez05}.  Although such results
are restricted for $\Gamma<2$ and $r_0=0$, the negative density for
small $r$ in the case $\Gamma\to2^{-}$ would take the
form~\cite{Tellez05}
\begin{eqnarray}
\rho_-(r)\displaystyle\mathop{=}_{r\to 0}
A_1r^{-2n}+O(r^{2(1-n)}), \qquad \text{for\ } n<1
\,,
\end{eqnarray}
where $A_1$ is a constant. The results shown in table
\ref{table1} match last predictions in the case $n<1$ and
$r_0\rightarrow0$, ie.~$\rho_{-}(r)\propto r^{-2n}$. When $n>1$, this
behavior changes, in part due to the collapse of the negative ions
into the charged impurity.
\begin{table}[h]
\centering
{\renewcommand{\arraystretch}{1.4}
\renewcommand{\tabcolsep}{0.075cm}
\begin{tabular}{|c|c|c|c|c|c|c|c|c|c|c|c|c|}
\hline
$mr_0$ & \multicolumn{12}{|c|}{$n$}  \\
\hline
& $7$ & $5.2$ & $4$ & $3.5$ & $2.7$ & $2$ & $1.4$ & $1$ & $0.8$ & $0.6$ & $0.4$ & $0.2$  \\
\hline\hline
$10^{-1}$ & $-8.92$ & $-7.25$ & $-5.88$  & $-5.33$ & $-4.37$ & $-3.43$& $-2.47$ & $-1.65$ & $-1.23$ & $-0.81$ & $-0.40$ & $-0.12$  \\
\hline
$10^{-3}$ & $-9.11$ & $-7.42$ & $-6.13$  & $-5.52$ & $-4.65$ & $-3.81$& $-2.80$ & $-2.00$ & $-1.60$ & $-1.20$ & $-0.80$ & $-0.40$  \\
\hline
$10^{-6}$ & $-9.15$ & $-7.44$ & $-6.20$  & $-5.53$ & $-4.66$ & $-3.90$& $-2.80$ & $-2.00$ & $-1.60$ & $-1.20$ & $-0.80$ & $-0.40$  \\
\hline
\end{tabular}}
\caption{Slopes of the plots $\ln(m^{-2}\rho_-(r))$ vs.~$\ln(r/r_0)$. The slope describes the linear behavior present in the
curves shown, for instance, in figure \ref{fig4}; we considered different values for the charge of the impurity $n$.
While the slope takes the value $\Gamma_n=-n\beta e^2=-2n$ for $n<1$ and $r_0\to0$, the slope does not satisfy this relation
as $n>1$ or $r_0$ is large enough. The larger $n$ the lesser the slope approximates to $-2n$. For $n=1$ the slope also satisfies this condition. Our results match the predictions stated in \cite{Tellez05} found when $\Gamma<2$ and $r_0\to0$. We used $m^{-2}\rho^0=0.87$ and $mR=30$.}
\label{table1}
\end{table}

\begin{table}[t]
\centering
{\renewcommand{\arraystretch}{1.4}
\renewcommand{\tabcolsep}{0.1cm}
\begin{tabular}{|c|c|c|c|c|c|c|c|c|c|}
\hline
$mr_0$ & \multicolumn{9}{|c|}{$n$}  \\
\hline
          & $7$     & $5.3$  & $3.5$  & $2$ & $1$ & $0.7$ & $0.5$ & $0.3$ & $0.1$ \\
\hline\hline
$10^{-1}$ & $6.51$  & $5.91$ & $5.01$ & $3.66$& $1.94$ &$1.15$ & $0.586$ & $3.76\!\times\!\!10^{-2}$ & $-0.464$ \\
\hline
$10^{-2}$ & $11.1$  & $10.5$ & $9.55$ & $8.16$& $5.82$ &$4.11$ & $2.75$ & $1.37$ & $7.11\!\times\!\!10^{-3}$ \\
\hline
$10^{-3}$ & $15.7$  & $15.1$ & $14.2$ & $12.7$& $10.0$ &$7.29$ & $5.07$ & $2.82$ & $0.607$  \\
\hline
$10^{-4}$ & $20.3$  & $19.7$ & $18.8$ & $17.3$& $14.4$ &$10.5$ & $7.37$ & $4.23$ & $1.22$ \\
\hline
$10^{-5}$ & $24.9$  & $24.3$ & $23.4$ & $21.9$& $18.7$ &$13.7$ & $9.67$ & $5.62$ & $1.80$ \\
\hline
$10^{-6}$ & $29.5$  & $28.9$ & $28.0$ & $26.6$& $23.2$ &$16.9$ & $12.0$ & $7.01$ & $2.34$ \\
\hline
$10^{-7}$ & $34.1$  & $33.5$ & $32.6$ & $31.1$& $27.6$ &$20.2$ & $14.3$ & $8.39$ & $2.87$ \\
\hline
$10^{-8}$ & $38.7$  & $38.1$ & $37.2$ & $35.7$& $32.1$ &$23.4$ & $16.6$ & $9.77$ & $3.37$ \\
\hline
$10^{-9}$ & $43.3$  & $42.8$ & $41.8$ & $40.3$& $36.6$ &$26.6$ & $18.9$ & $11.2$ & $3.87$ \\
\hline
$10^{-10}$ & $48.0$  & $47.4$ & $46.4$ & $44.9$& $41.1$ &$29.8$ & $21.2$ & $12.5$ & $4.35$ \\
\hline
\end{tabular}}
\caption{Value of $\ln(m^{-2}\rho_-(r))$ at $r=r_0$ for different values of $r_0$ and $n$.
As expected, the smaller $r_0$, the larger the negative density at $r=r_0$ because the collapse
of the counter-ions is stronger. For $n>1$ the values of the negative density at $r_0$ take a very
large value as $r_0\to0$. For $n<1$ such values are also large but notably decrease
because of the decrease of negative ions already described. We used $m^{-2}\rho^0=0.87$ and $mR=30$.}
\label{table2}
\end{table}

\subsection{Asymptotic limits}

It is interesting to study the limiting cases in which
$r_0\rightarrow0$ and $R\rightarrow\infty$. When $r_0\rightarrow0$ we
have that
$[t_{l+\nu}^{(0)}]^{-1}\sim\frac{2}{\Gamma(l+\nu)\Gamma(l+\nu+1)}\left(\frac{r_0}{2}\right)^{2(l+\nu)}$
for $l>0$ and $[t_{0}^{(0)}]^{-1}\sim\frac{2(1-\delta_{0\nu})}{\Gamma(\nu)\Gamma(\nu+1)}\left(\frac{r_0}{2}\right)^{2\nu}
+\frac{m^2}{2\pi}\frac{\delta_{0\nu}}{\rho^0(r_0)}$, where $\rho^0(r_0)=\rho^0(a=r_0)$, for $l=0$.
Although such terms vanish as $r_0\to0$, a term of the form
$[t_{l+\nu}^{(0)}]^{-1}[K_{l+\nu}^{(r)}]^2$ does not necessarily cancel for
small values of $r$ because $K_{l+\nu}^{(r)}$ diverges.
These contributions can be neglected, however, when $r\gg r_0$;
in this limit we have
\begin{align}\label{densi1}
\frac{2\pi}{m^2}\rho_{+}^{(\nu_+,0,\tilde R)}(\mathbf r)
&=
\frac{2\pi}{m^2}\tilde \rho^0_{\nu}
+\sum_{l=1}^{\infty}
t_{l-\nu-1}^{(R)}[\,I_{l-\nu}^{(r)}]^{2}
+\sum_{l=1}^{k-1}
t_{l+\nu+1}^{(R)}[\,I_{l+\nu}^{(r)}]^{2}
-\sum_{l=k}^{\infty}
t_{l+\nu}^{(R)}[\,I_{l+\nu}^{(r)}]^{2}
\nonumber\\
&
-(1-\delta_{0k})\frac{\frac{m^2}{2\pi}\frac{\delta_{0\nu}}{\rho^0(r_0)}[K_0^{(r)}]^2}
{1+\frac{m^2}{2\pi}\frac{\delta_{0\nu}}{\rho^0(r_0)}t_1^{(R)}}
-(1-\delta_{0k})t_{1+\nu}^{(R)}\frac{\frac{2m^2}{2\pi}\frac{\delta_{0\nu}}{\rho^0(r_0)}I_0^{(r)}K_0^{(r)}-[I_\nu^{(r)}]^2}
{1+\frac{m^2}{2\pi}\frac{\delta_{0\nu}}{\rho^0(r_0)}t_1^{(R)}},
\\
\frac{2\pi}{m^2}\rho_{-}^{(\nu_+,0,\tilde R)}(\mathbf r)&=
\frac{2\pi}{m^2}\tilde \rho^0_{\nu}
-\sum_{l=1}^{\infty}
t_{l-\nu}^{(R)}[\,I_{l-\nu}^{(r)}]^{2}
-\sum_{l=2}^{k}
t_{l+\nu}^{(R)}[\,I_{l+\nu}^{(r)}]^{2}
+\sum_{l=k+1}^{\infty}
t_{l+\nu-1}^{(R)}[\,I_{l+\nu}^{(r)}]^{2}
\nonumber\\
&
-t_\nu^{(R)}\frac{2d_{\,\nu}^{\,0}I_\nu^{(r)}K_\nu^{(r)}+[I_\nu^{(r)}]^2}{1+d_{\,\nu}^{\,0}t_\nu^{(R)}}
-(1-\delta_{0k})t_{1+\nu}^{(R)}\frac{\frac{2m^2}{2\pi}\frac{\delta_{0\nu}}{\rho^0(r_0)}I_1^{(r)}K_1^{(r)}+[I_{1+\nu}^{(r)}]^2}
{1+\frac{m^2}{2\pi}\frac{\delta_{0\nu}}{\rho^0(r_0)}t_1^{(R)}}
\nonumber\\
&
+\frac{d_{\,\nu}^{\,0}[K_{\nu}^{(r)}]^{2}}
{1+d_{\,\nu}^{\,0}t_{\nu}^{(R)}}
+(1-\delta_{0k})\frac{\frac{m^2}{2\pi}\frac{\delta_{0\nu}}{\rho^0(r_0)}[K_1^{(r)}]^2}
{1+\frac{m^2}{2\pi}\frac{\delta_{0\nu}}{\rho^0(r_0)}t_1^{(R)}}.
\end{align}
Another interesting limit is the case $R\rightarrow\infty$. In this limit $t_{l+\nu}^{(R)}\rightarrow0$ but
terms of the form $t_l^{(R)}[I_l^{(r)}]^2$ do not necessarily vanish as $r\rightarrow\infty$.
However, if we consider the case $r\ll R$ we find
\begin{eqnarray}\label{densi2}
\frac{2\pi}{m^2}\rho_{+}^{(\nu_+,\tilde r_0,\infty)}(\mathbf r)\!\!\!&=&\!\!\!
\frac{2\pi}{m^2}\tilde \rho^0_{\nu}
-\sum_{l=1}^{\infty}[t_{l-\nu}^{(0)}]^{-1}[K_{l-\nu}^{(r)}]^2
-\sum_{l=0}^{k-1}[t_{l+\nu}^{(0)}]^{-1}[K_{l+\nu}^{(r)}]^2
\nonumber\\
\!\!\!&&\!\!\!
+\sum_{l=k}^{\infty}[t_{l+\nu+1}^{(0)}]^{-1}[K_{l+\nu}^{(r)}]^{2},
\\
\frac{2\pi}{m^2}\rho_{-}^{(\nu_+,\tilde r_0,\infty)}(\mathbf r)\!\!\!&=&\!\!\!
\frac{2\pi}{m^2}\tilde \rho^0_{\nu}
+\sum_{l=1}^{\infty}[t_{l-\nu+1}^{(0)}]^{-1}[K_{l-\nu}^{(r)}]^{2}
+\sum_{l=0}^{k}[t_{l+\nu-1}^{(0)}]^{-1}[K_{l+\nu}^{(r)}]^{2}
\nonumber\\
\!\!\!&&\!\!\!
-\sum_{l=k+1}^{\infty}[t_{l+\nu}^{(0)}]^{-1}[K_{l+\nu}^{(r)}]^{2}.
\end{eqnarray}
Finally, we can combine both limits $r_0\rightarrow0$ and $R\rightarrow\infty$. In the interval $r_0\ll r\ll R$ we get
\begin{align}\label{densi3}
\rho_{+}^{(\nu_+,0,\infty)}(\mathbf r)
&=
\tilde \rho^0_{\nu}
-\left(\frac{m^2}{2\pi}\right)^2\frac{(1-\delta_{0k})\delta_{0\nu}}{\rho^0(r_0)}[K_0^{(r)}]^2,
\\\label{densi3a}
\rho_{-}^{(\nu_+,0,\infty)}(\mathbf r)
&=
\tilde \rho^0_{\nu}
+\left(\frac{m^2}{2\pi}\right)^2\frac{(1-\delta_{0k})\delta_{0\nu}}{\rho^0(r_0)}[K_1^{(r)}]^2
+\frac{m^2}{2\pi}d_{\,\nu}^{\,0}[K_{\nu}^{(r)}]^{2}.
\end{align}
From Eqs. (\ref{densi3}) and (\ref{densi3a}) we can obtain the particular cases
\begin{align}\label{densi4}
\rho_{+}^{(\nu_+,0,\infty)}(\mathbf r)
&=\left\{\begin{array}{ll}
\rho^0, & n=0\\
\frac{\rho_{++}^{(2)0}(r,0)}{\rho^0(r_0)}, & n\ge1, n\in\mathbb{Z}\\
\tilde \rho^0_{\nu}, & n\notin\mathbb{Z}
\end{array}\right.
\\
\rho_{-}^{(\nu_+,0,\infty)}(\mathbf r)
&=\left\{\begin{array}{ll}
\rho^0, & n=0\\
\frac{\rho_{+-}^{(2)0}(r,0)}{\rho^0(r_0)}, & n\ge1, n\in\mathbb{Z}\\
\tilde \rho^0_{\nu}+\frac{m^2}{2\pi}d_{\,\nu}^{\,0}[K_{\nu}^{(r)}]^{2}, & n\notin\mathbb{Z}
\end{array}\right.
\end{align}
where $\rho_{+\pm}^{(2)0}(r,0)$ are the two-point correlation
functions for the unperturbed plasma \cite{ref-corn}.  Notice that for
$n\ge1$ and $n\in\mathbb{Z}$ both the positive and negative densities
do not depend on $n$ (their respective densities are the same for any
integer charge). This takes place because an amount of $k-1$
counter-ions screen the impurity as their collapse cannot be avoided
in the limit $r_0\to0$ leaving a charge $n=1$ unscreened.

\section{Integrated Charge}\label{char}

We now focus on finding the integrated charge of the system contained
in an annulus region from $r_0$ up to a distance $r>r_0$ . Let
$e_{\pm}(r,r_0)=\int_{r_0}^{r}\int_0^{2\pi}\rho_{\pm}(\mathbf
r)d^{\,2}r$. We now define the following functions
\begin{eqnarray}\label{eq-func}
a_j(r,r_0)\!\!\!&=&\!\!\!\frac{m^2}{2\pi}\int_{r_0}^r\int_0^{2\pi}\!\!d^2r'[K_j^{(r')}]^2
=\frac{x^2}{2}\Big[[\,K_j^{(x)}]^2-K_{j-1}^{(x)}K_{j+1}^{(x)}\Big]\Bigg\vert^{\tilde r}_{\tilde r_0},
\\
b_j(r,r_0)\!\!\!&=&\!\!\!\frac{m^2}{2\pi}\int_{r_0}^r\int_{0}^{2\pi}\!\!d^2r'[\,I_j^{(r')}]^2
=\frac{x^2}{2}\Big[[I_j^{(x)}]^2-I_{j-1}^{(x)}I_{j+1}^{(x)}\Big]\Bigg\vert^{\tilde r}_{\tilde r_0},
\\
c_j(r,r_0)\!\!\!&=&\!\!\!\frac{m^2}{2\pi}\int_{r_0}^r\int_{0}^{2\pi}\!\!d^2r'I_j^{(r')}K_j^{(r')},
\\\label{eq-func3}
e_{0}(r,r_0)\!\!\!&=&\!\!\!\int_{r_0}^{r}\int_0^{2\pi} d^2r'\rho^{0}=
\frac{\tilde r^2-\tilde r_0^2}{2}\left(\,\ln\frac{2}{ma}-\gamma\right).
\end{eqnarray}
The total positive and negative charges in the annulus region can
easily be found by integrating the densities found in the previous
section and using Eqs.~(\ref{eq-func})--(\ref{eq-func3}). The
integrated charge, from $r_0$ up to a distance $r$, is given by
$e_{t}(r,r_0)=\int_{r_0}^{r}\int_0^{2\pi}d^{\,2}r'(\rho_{+}(r')-\rho_{-}(r'))$.
Its general expression is given by
\begin{align}\label{et-gen}
e_t^{(\nu_+,\tilde r_0,\tilde R)}(r)=
&-\sum_{l=0}^{k-1}\frac{[t_{l+\nu}^{(0)}]^{-1}A_{l+\nu}(r,r_0)}{1+[t_{l+\nu}^{(0)}]^{-1}t_{l+\nu+1}^{(R)}}
-\sum_{l=0}^{k-1}t_{l+\nu+1}^{(R)}\frac{2[t_{l+\nu}^{(0)}]^{-1}C_{l+\nu}(r,r_0)-B_{l+\nu}(r,r_0)}{1+[t_{l+\nu}^{(0)}]^{-1}t_{l+\nu+1}^{(R)}}
\nonumber\\
&-\sum_{l=1}^{\infty}\frac{[t_{l-\nu+1}^{(0)}]^{-1}A_{l-\nu}(r,r_0)}{1+[t_{l-\nu+1}^{(0)}]^{-1}t_{l-\nu}^{(R)}}
+\sum_{l=1}^{\infty}t_{l-\nu}^{(R)}\frac{2[t_{l-\nu+1}^{(0)}]^{-1}C_{l-\nu}(r,r_0)+B_{l-\nu}(r,r_0)}{1+[t_{l-\nu+1}^{(0)}]^{-1}t_{l-\nu}^{(R)}}
\nonumber\\
&+\sum_{l=k}^{\infty}\frac{[t_{l+\nu+1}^{(0)}]^{-1}A_{l+\nu}(r,r_0)}{1+[t_{l+\nu+1}^{(0)}]^{-1}t_{l+\nu}^{(R)}}
-\sum_{l=k}^{\infty}t_{l+\nu}^{(R)}\frac{2[t_{l+\nu+1}^{(0)}]^{-1}C_{l+\nu}(r,r_0)+B_{l+\nu}(r,r_0)}{1+[t_{l+\nu+1}^{(0)}]^{-1}t_{l+\nu}^{(R)}}
\nonumber\\
&-\frac{[t_{\nu-1}^{(0)}]^{-1}a_{\nu}(r,r_0)}{1+[t_{\nu-1}^{(0)}]^{-1}t_{\nu}^{(R)}}
+t_{\nu}^{(R)}\frac{2[t_{\nu-1}^{(0)}]^{-1}c_{\nu}(r,r_0)+b_{\nu}(r,r_0)}{1+[t_{\nu-1}^{(0)}]^{-1}t_{\nu}^{(R)}}
\nonumber\\
&-\frac{[t_{1-\nu}^{(0)}]^{-1}a_{1-\nu}(r,r_0)}{1+[t_{1-\nu}^{(0)}]^{-1}t_{-\nu}^{(R)}}
-t_{-\nu}^{(R)}\frac{2[t_{1-\nu}^{(0)}]^{-1}c_{1-\nu}(r,r_0)-b_{1-\nu}(r,r_0)}{1+[t_{1-\nu}^{(0)}]^{-1}t_{-\nu}^{(R)}}
\end{align}
where we used the definitions
\begin{eqnarray}
a_{l+\nu}(r,r_0)+a_{l+\nu+1}(r,r_0)\!\!\!&=&\!\!\!
-xK_{l+\nu}^{(x)}K_{l+\nu+1}^{(x)}\Big\vert_{r_0}^{r}
\equiv A_{l+\nu}(r,r_0),
\\
b_{l+\nu}(r,r_0)+b_{l+\nu+1}(r,r_0)\!\!\!&=&\!\!\!
xI_{l+\nu}^{(x)}I_{l+\nu+1}^{(x)}\Big\vert_{r_0}^{r}
\equiv B_{l+\nu}(r,r_0),
\\
c_{l+\nu}(r,r_0)-c_{l+\nu+1}(r,r_0)\!\!\!&\equiv&\!\!\!
C_{l+\nu}(r,r_0).
\end{eqnarray}

\subsection{Integrated Charge at the boundary}

An interesting fact is evaluating the total integrated charge at the
boundary of the system, i.e., at $r=R$. We should expect this charge
to vanish because, excluding the impurity, the system is neutral. We
can easily check that in the limits $\tilde R\gg1$ and $\tilde
r_0\ll1$ we have
\begin{eqnarray}
A_{l+\nu}(R,r_0)\!\!\!&=&\!\!\!
\tilde r_0 K_{l+\nu+1}(\tilde r_0)K_{l+\nu}(\tilde r_0)
+O(e^{-\tilde R}),
\\
B_{l+\nu}(R,r_0)\!\!\!&=&\!\!\!
\tilde R I_{l+\nu+1}(\tilde R)I_{l+\nu}(\tilde R)
+O(\tilde r_0^{2(l+\nu+1)}).
\end{eqnarray}
Since $t_{l+\nu+1}^{(R)}\sim e^{-2\tilde R}$, the contributions coming from the terms $C_l(R,r_0)$ are exponentially small and so can be neglected. We thus have
\begin{eqnarray}
e_{t}^{(\nu_+,\tilde r_0,\tilde R)}(R)\!\!\!&=&\!\!\!
-\sum_{l=1}^{\infty}\frac{\tilde r_0 I_{l-\nu+1}^{(0)}K_{l-\nu}^{(0)}}
{1+[t_{l-\nu+1}^{(0)}]^{-1}t_{l-\nu}^{(R)}}
-\sum_{l=0}^{k-1}\frac{\tilde r_0I_{l+\nu}^{(0)}K_{l+\nu+1}^{(0)}}
{1+[t_{l+\nu}^{(0)}]^{-1}t_{l+\nu+1}^{(R)}}
+\sum_{l=k}^{\infty}\frac{\tilde r_0I_{l+\nu+1}^{(0)}K_{l+\nu}^{(0)}}
{1+[t_{l+\nu+1}^{(0)}]^{-1}t_{l+\nu}^{(R)}}
\nonumber\\
\!\!\!&&\!\!\!
+\sum_{l=1}^{\infty}\frac{\tilde R K_{l-\nu}^{(R)}I_{l-\nu+1}^{(R)}}
{1+[t_{l-\nu+1}^{(0)}]^{-1}t_{l-\nu}^{(R)}}
+\sum_{l=0}^{k-1}\frac{\tilde R K_{l+\nu+1}^{(R)}I_{l+\nu}^{(R)}}
{1+[t_{l+\nu}^{(0)}]^{-1}t_{l+\nu+1}^{(R)}}
-\sum_{l=k}^{\infty}\frac{\tilde R K_{l+\nu}^{(R)}I_{l+\nu+1}^{(R)}}
{1+[t_{l+\nu+1}^{(0)}]^{-1}t_{l+\nu}^{(R)}}\,\,\,\,\,
\nonumber\\
\!\!\!&&\!\!\!
-\frac{\tilde r_0I_{1-\nu}^{(0)}K_{\nu}^{(0)}+d_\nu^{0}a_{\nu}(R,r_0)}
{1+t_{\nu}^{(R)}(d_{\nu}^{\,0}+[t_{1-\nu}^{(0)}]^{-1})}
+\frac{\tilde R K_{\nu}^{(R)}I_{1-\nu}^{(R)}}
{1+t_{\nu}^{(R)}(d_\nu^{\,0}+[t_{1-\nu}^{(0)}]^{-1})}
+O(e^{-\tilde R},\tilde r_0^{2(l+1)}).
\end{eqnarray}
In the same approximation all the denominators behave like $\sim\frac{1}{1+O(e^{-\tilde R})}=1+O(e^{-\tilde R})$. So
\begin{eqnarray}
e_{t}^{(\nu_+,\tilde r_0,\tilde R)}(R)\!\!\!&=&\!\!\!
-\tilde r_0\sum_{l=0}^{k-1}\Big[I_{l-\nu+1}^{(0)}K_{l-\nu}^{(0)}+I_{l+\nu}^{(0)}K_{l+\nu+1}^{(0)}\Big]
+\tilde r_0\sum_{l=k}^{\infty}\Big[I_{l+\nu+1}^{(0)}K_{l+\nu}^{(0)}-I_{l-\nu+1}^{(0)}K_{l-\nu}^{(0)}\Big]
\nonumber\\
\!\!\!&&\!\!\!
+\tilde R \sum_{l=0}^{k-1}\Big[K_{l-\nu}^{(R)}I_{l-\nu+1}^{(R)}+K_{l+\nu+1}^{(R)}I_{l+\nu}^{(R)}\Big]
-\tilde R \sum_{l=k}^{\infty}\Big[K_{l+\nu}^{(R)}I_{l+\nu+1}^{(R)}-K_{l-\nu}^{(R)}I_{l-\nu+1}^{(R)}\Big]
\nonumber\\
\!\!\!&&\!\!\!
-d_\nu^{\,0}a_{\nu}(R,r_0)+O(e^{-\tilde R},\tilde r_0^{2(l+1)}).
\end{eqnarray}
When $\nu=0$, it is easy to see, using the Wronskian relations for the
Bessel functions~\cite{Grad}, that $e_{t^+}^{(0,\tilde r_0,\tilde
  R)}(R)=0$, so the total charge of the system is $ne$ (the charge of
the impurity). This means that the charge distribution generated by
the impurity close to the origin is compensated by a charge
distribution of opposite sign in the boundary. For $\nu\neq0$ the
charge at $r=R$ does not simplify and so numerical solutions must be
found; in this case we also find $e_{t}^{(\nu,\tilde r_0,\tilde
  R)}(R)=0$.

\begin{figure}
\centering
\includegraphics[scale=0.85]{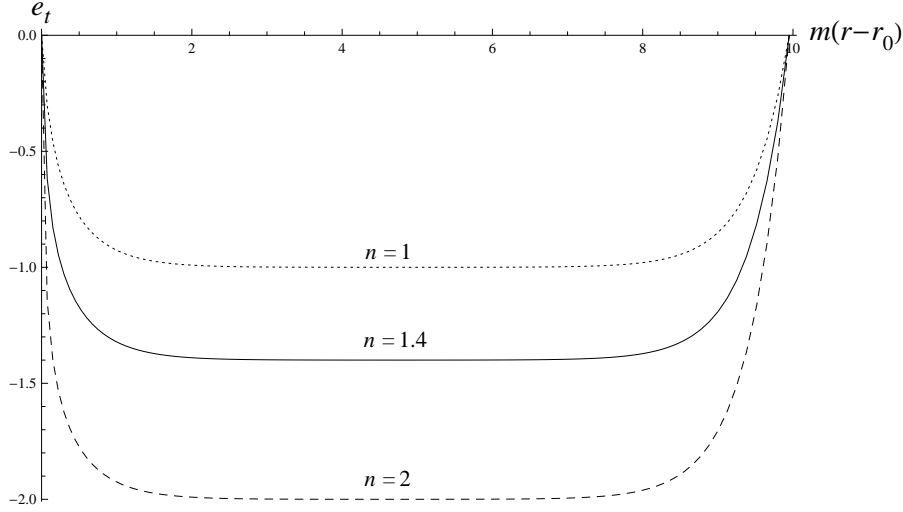}
\caption{Integrated charge in the plasma for three different impurities located at the origin. The size of the impurity is $mr_0=0.05$ and the size of the disk is $mR=10$. }
\label{fig5}
\end{figure}

\begin{figure}
\centering
\includegraphics[scale=0.85]{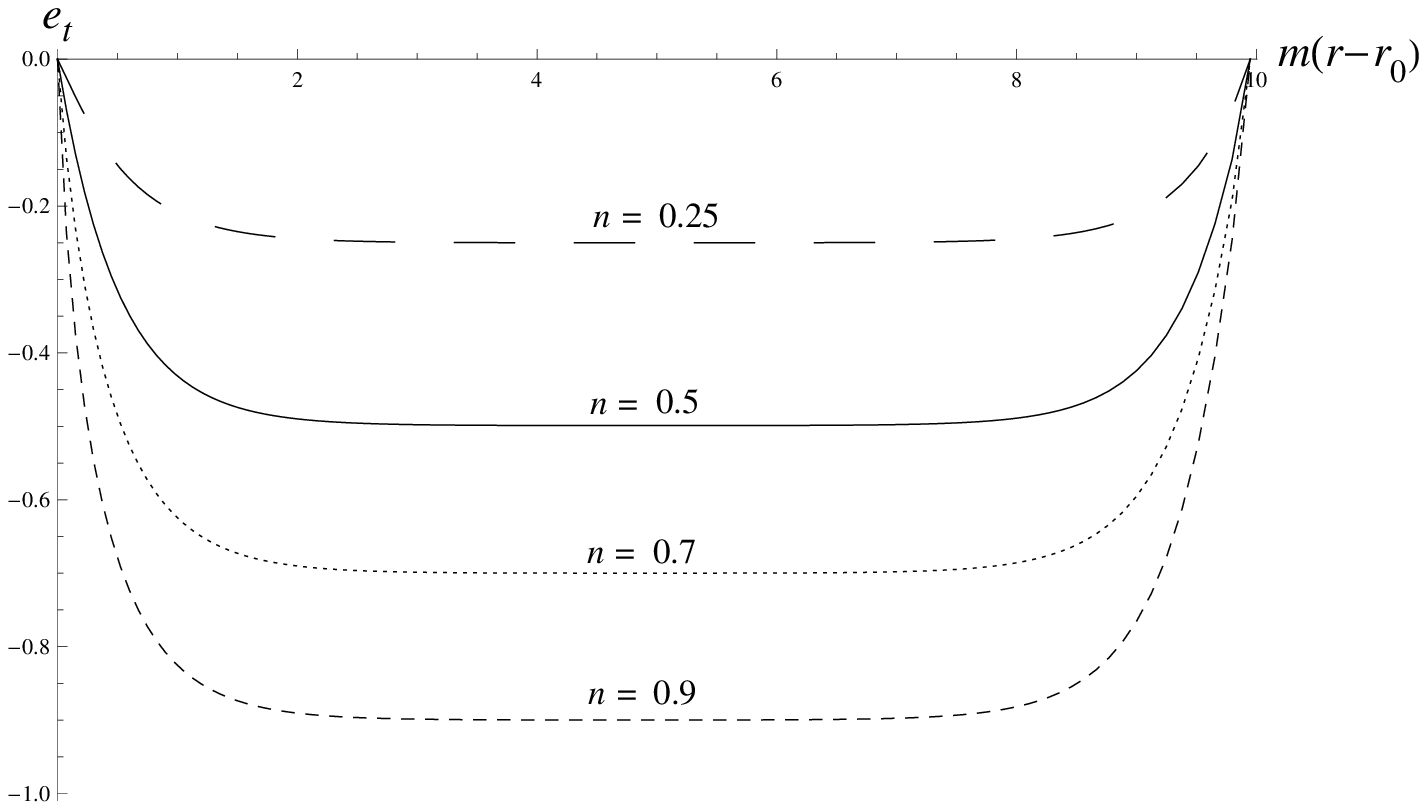}
\caption{Integrated charge in the plasma for four different impurities located at the origin. The size of the impurity is $mr_0=0.05$ and the size of the disk is $mR=10$. }
\label{fig6}
\end{figure}

Figures \ref{fig5} and \ref{fig6} show the integrated charge for
different values of the impurity charge $n$. By definition it starts
with a zero value at $r=r_0\neq0$. Then it decreases to reach a valley with a
value approaching $-n$ as expected since the plasma screens the
impurity. Then, close the outer boundary ($r\to R$), the integrated
charge increases to reach the zero value, as discussed above.


\subsection{Asymptotic Limits}
We will evaluate the same asymptotic cases than those analyzed in
section \ref{dens}.  For $r_0\rightarrow 0$ and $r\gg r_0$ the net
density becomes
\begin{align}\label{et}
e_t^{(\nu_+,0,\tilde R)}(r)=
&-k+\delta_{0\nu}\!\left(1-\frac{1}{1+\frac{m^2}{2\pi}\frac{\delta_{0\nu}}{\rho^0(r_0)}t_{1+\nu}^{(R)}}\right)\!
-(1-\delta_{0k})t_{1+\nu}^{(R)}\frac{\frac{2m^2}{2\pi}\frac{\delta_{0\nu}}{\rho^0(r_0)}C_{0}(r,0)-B_{\nu}(r,0)}
{1+\frac{m^2}{2\pi}\frac{\delta_{0\nu}}{\rho^0(r_0)}t_{1+\nu}^{(R)}}
\nonumber\\
&+r\sum_{l=1}^{k-1}t_{l+\nu+1}^{(R)}I_{l+\nu+1}^{(r)}I_{l+\nu}^{(r)}
+r\sum_{l=1}^{\infty}t_{l-\nu}^{(R)}I_{l-\nu+1}^{(r)}I_{l-\nu}^{(r)}
-r\sum_{l=k}^{\infty}t_{l+\nu}^{(R)}I_{l+\nu+1}^{(r)}I_{l+\nu}^{(r)}
\nonumber\\
&-\frac{d_{\,\nu}^{\,0}a_{\nu}(r,0)}{1+d_{\,\nu}^{\,0}t_{\nu}^{(R)}}
+t_{\nu}^{(R)}\frac{2d_{\,\nu}^{\,0}c_{\nu}(r,0)+b_{\nu}(r,0)+b_{1-\nu}(r,0)}{1+d_{\,\nu}^{\,0}t_{\nu}^{(R)}}
\end{align}
When $R\rightarrow\infty$ and $r\ll R$ we have
\begin{align}\label{et2}
e_t^{(\nu_+,\tilde r_0,\infty)}(r)=
&-\sum_{l=0}^{k-1}[t_{l+\nu}^{(0)}]^{-1}A_{l+\nu}(r,r_0)
-\sum_{l=1}^{\infty}[t_{l-\nu+1}^{(0)}]^{-1}A_{l-\nu}(r,r_0)
\nonumber\\
&+\sum_{l=k}^{\infty}[t_{l+\nu+1}^{(0)}]^{-1}A_{l+\nu}(r,r_0)
-[t_{\nu-1}^{(0)}]^{-1}a_{\nu}(r,r_0)
-[t_{1-\nu}^{(0)}]^{-1}a_{1-\nu}(r,r_0)
\end{align}
In the combined limits $r_0\rightarrow0$ and $R\rightarrow\infty$, we get, in the interval $r_0\ll r\ll R$
\begin{align}\label{carga3}
e_t^{(\nu_+,0,\infty)}(r)=
&-k-d_{\,\nu}^{\,0}a_{\nu}(r,0)
\end{align}

\begin{figure}
\centering
\includegraphics[scale=0.85]{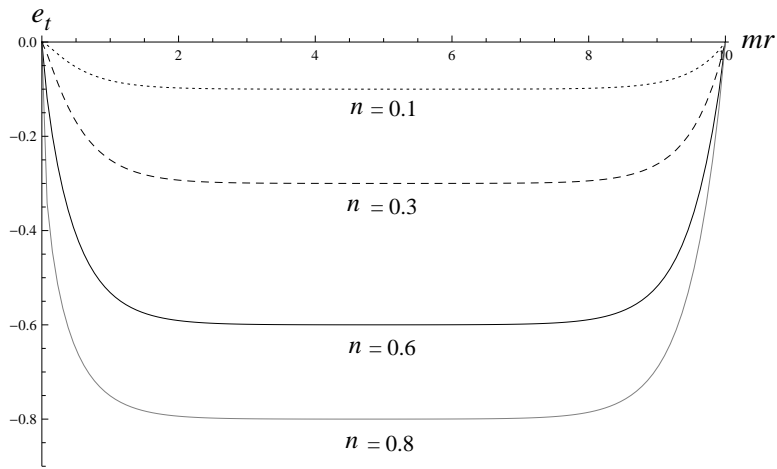}
\caption{Integrated charge in the plasma for four different fractional
  impurities located at the origin. The impurity is a treated as point
  particle ($r_0=0$) according to Eq.~(\ref{et}). The size of the disk
  is $mR=10$.}
\label{fig7}
\end{figure}

\begin{figure}
\centering
\includegraphics[scale=0.85]{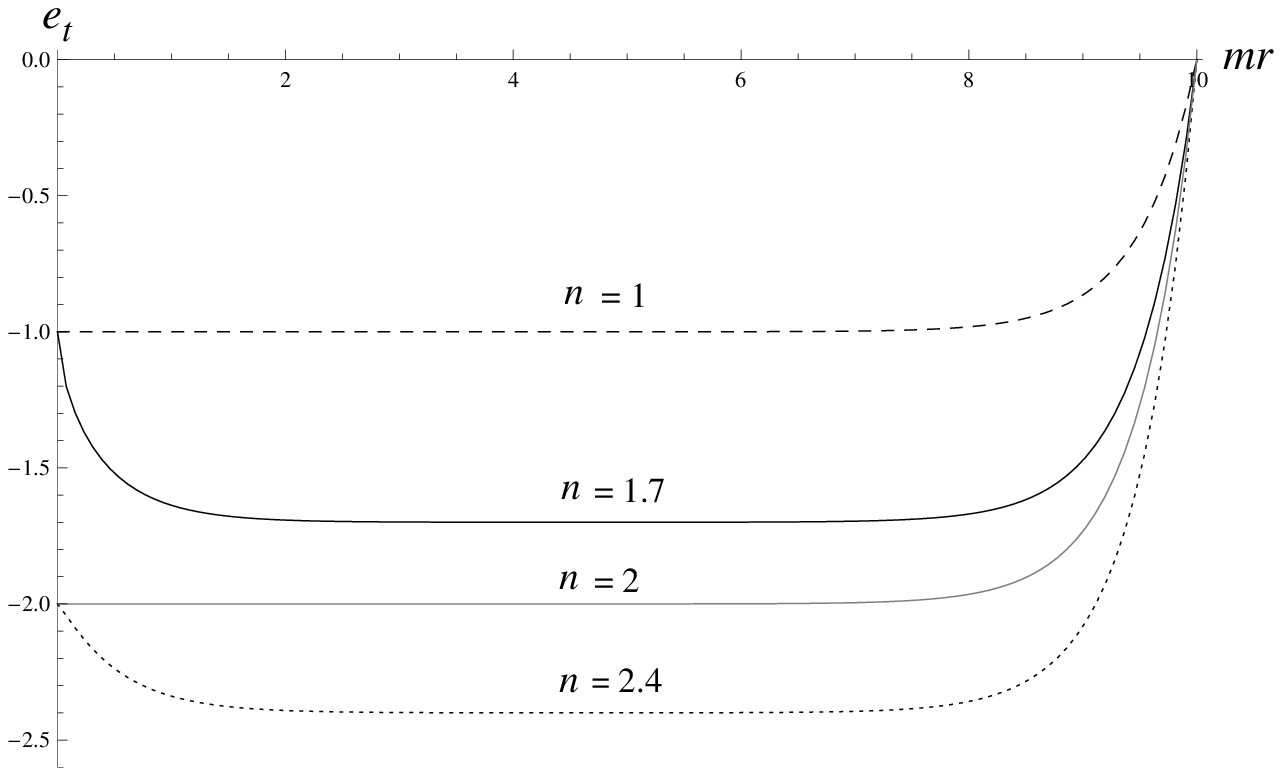}
\caption{Integrated charge in the plasma for four different impurities
  located at the origin. The impurity is a treated as point particle
  ($r_0=0$) according to Eq.~(\ref{et}). The size of the disk is
  $mR=10$. }
\label{fig8}
\end{figure}

Figs.~\ref{fig7} and \ref{fig8} show the integrated charge in the case
when $r_0=0$. Fig.~\ref{fig7} is, in essence, similar to
Fig.~\ref{fig6} when $r_0\neq 0$. However, Fig.~\ref{fig8} shows a new
interesting effect, which is not present in the case when $r_0\neq 0$
(Fig.~\ref{fig5}). When $n$ is an integer, the accumulated charge
$e_t(r)$ starts at $r=r_0=0$ with a value equal to $-n$, instead of
zero, as it was the case when $r_0\neq 0$. This means that exactly $n$
ions of charge $-e$ of the plasma (counter-ions) have collapsed at the
position $r_0=0$ where the impurity is, and they have completely
neutralized it. On the other hand, when $n=k+\nu$ is not an integer
($\nu\neq 0$), $k$ negative ions will partially neutralize the
impurity by collapsing into it at position $r_0=0$, but a fractional 
charge $e \nu$ remains to be screened, which cannot be completely
neutralized by collapsing into it, because the charges of ions the
plasma are integers. Nevertheless, this remaining charge is screened
by a difuse layer of plasma ions, of tipical length given by the
screening length $m^{-1}$. This situation is illustrated with the
cases $n=1.7$ and $n=2.4$ in Fig.~\ref{fig8}.

\begin{figure}
\centering
\includegraphics[scale=0.85]{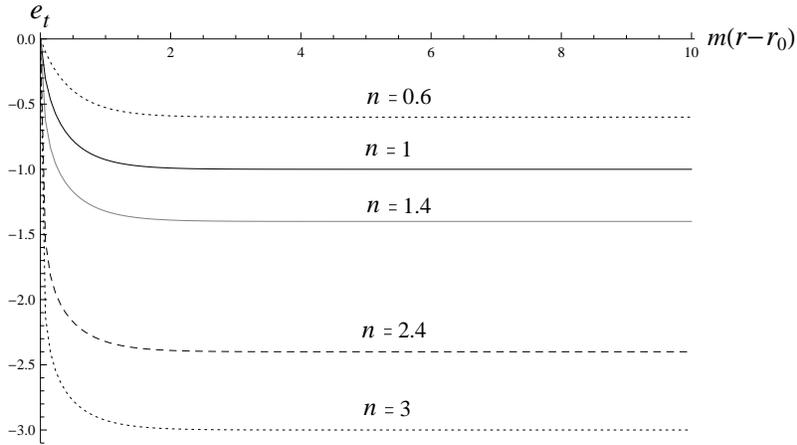}
\caption{Integrated charge in the plasma for five different impurities
  located at the origin. The impurity has a size $mr_0=0.05$ and we
  consider the limit $R\to\infty$ according to Eq.~(\ref{et2}). The
  density of the gas is constant with value $-n$ for large enough
  distances.}
\label{fig9}
\end{figure}

\begin{figure}
\centering
\includegraphics[scale=0.85]{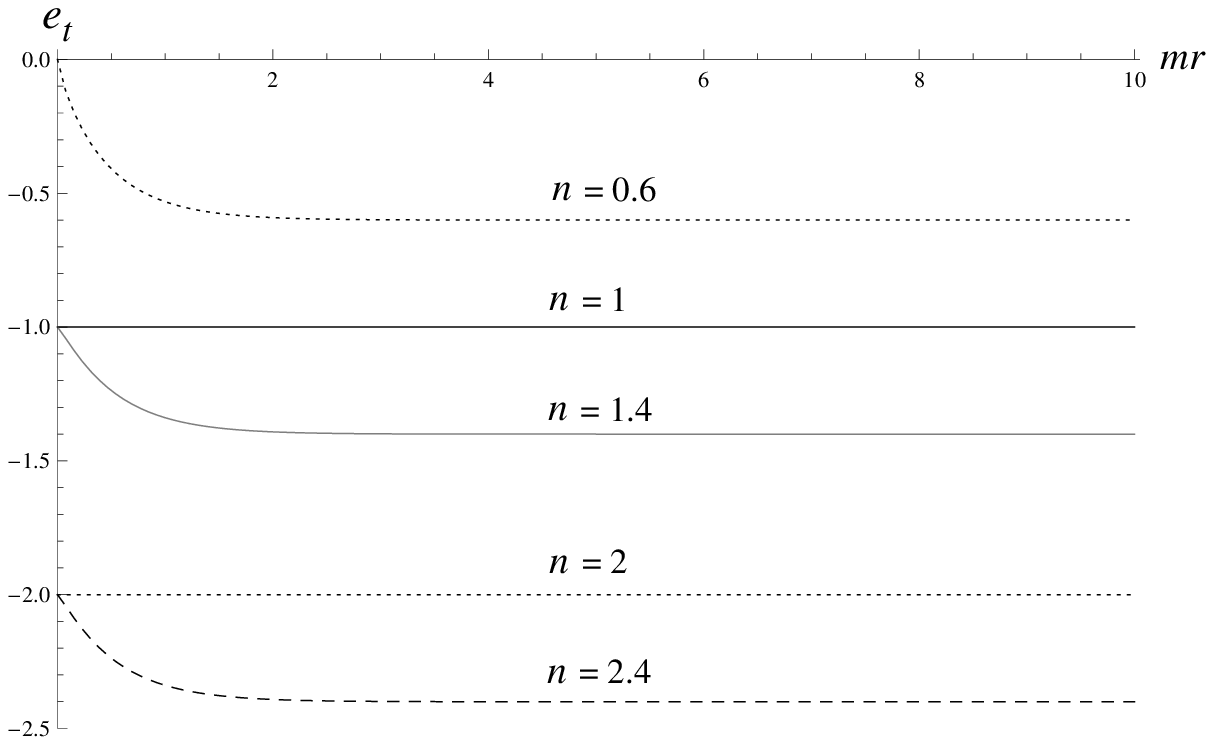}
\caption{Integrated charge in the plasma for five different impurities
  located at the origin. We consider the limits $r_0\to0$ and
  $R\to\infty$ according to Eq.~(\ref{carga3}). }
\label{fig10}
\end{figure}

Fig.~\ref{fig9} shows the integrated charge for $r_0\neq 0$ and
$R\to\infty$. Since the boundary of the large disk $R$ is now at
infinity, the difuse layer of remaining charge that usually
accumulates close that boundary, observed in the previous figures, is
now receded to infinity with the boundary and it cannot be observed
anymore in Fig.~\ref{fig9}. Fig.~\ref{fig10} shows the combined
effects of taking $r_0=0$ and $R\to\infty$.

\section{Grand Potential and Pressure}\label{poten}

In this section we will calculate the grand potential and the
partition function for the system we have discussed. Using
Eq.~(\ref{eq:pot1}) we see that the eigenvectors that generate the
partition function satisfy the equations
\begin{eqnarray}\label{eq:diff-vec}
m_+(\mathbf r)g(\mathbf r)\!\!\!&=&\!\!\!2\lambda\partial_zf(\mathbf r),
\nonumber\\
m_-(\mathbf r)f(\mathbf r)\!\!\!&=&\!\!\!2\lambda\partial_{\bar z}g(\mathbf r).
\end{eqnarray}
Remember that for this problem $m_\pm(\mathbf r)=m(\mathbf r)\left(\frac{r}{L}\right)^{\pm2n}$, where
\begin{eqnarray}\label{eq:fugac}
m(\mathbf r)&=&\left\{
\begin{array}{c}0,\,\,\,r<r_0\\m,\,\,\,r>r_0\end{array}\right..
\end{eqnarray}
In the region $r_0<r<R$ it is more convenient to define the functions
$\bar g(\mathbf r)=e^{-V(\mathbf r)}g(\mathbf r)$ and $\bar f(\mathbf r)=e^{V(\mathbf r)}f(\mathbf r)$.
Decomposing $\bar g(\mathbf r)$ as a Fourier series of the form $\bar g_l(\mathbf r)=\bar g_l(r)e^{il\theta}$, we find each mode satisfies the differential equation
\begin{eqnarray}\label{eq:diffge}
\Bigg\{\rho^2\frac{d^2}{d\rho^2}+\rho \frac{d}{d\rho}
-\Big[(l+n)^2+\rho^2\Big]\Bigg\}\bar g_l(\mathbf{\rho})=0,
\end{eqnarray}
with $\rho\equiv\frac{\tilde r}{\lambda}$. The general solutions are given by
\begin{eqnarray}\label{eq:sol-func}
g_l(\mathbf r)\!\!\!&=&\!\!\!\left(\frac{r}{L}\right)^{-n}e^{il\theta}\Big(B_lI_{l+n}(\rho)+A_lK_{l+n}(\rho)\Big),\nonumber\\
f_l(\mathbf r)\!\!\!&=&\!\!\!\left(\frac{r}{L}\right)^{n}e^{i(l+1)\theta}\Big(B_lI_{l+n+1}(\rho)-A_lK_{l+n+1}(\rho)\Big).
\end{eqnarray}
In the first region ($r<r_0$) and the third region ($r>R$), we have the general solutions
\begin{eqnarray}\label{eq:sol-func1}
g_l^{(1,3)}(\mathbf r)\!\!\!&=&\!\!\!u_l^{(1,3)}e^{il\theta}\left(\frac{_{mL}}{^{\lambda}}\right)^{n}\left(\frac{_{\tilde r}}{^{\lambda}}\right)^{l},\\
f_l^{(1,3)}(\mathbf r)\!\!\!&=&\!\!\!v_l^{(1,3)}e^{i(l+1)\theta}\left(\frac{_\lambda}{^{mL}}\right)^{n}\left(\frac{_{\tilde r}}{^{\lambda}}\right)^{-l-1}.
\end{eqnarray}
(The coefficients 1, 3 refer to the first and third regions.) The boundary conditions imply again that both $g_l^{(1)}(\mathbf r)$ and $f_l^{(1)}(\mathbf r)$ must be finite at $r=0$ while $g_l^{(3)}(\mathbf r)$ and $f_l^{(3)}(\mathbf r)$ must vanish at $r\rightarrow\infty$.
This generates the conditions
\begin{eqnarray}\label{eq:condiciones}
\left(\begin{array}{cc}
-K_{l+n+1}(\tilde r_0/\lambda) & I_{l+n+1}(\tilde r_0/\lambda)\\
K_{l+n}(\tilde R/\lambda) & I_{l+n}(\tilde R/\lambda)
\end{array}
\right)
\left(\begin{array}{c}
A_l\\ B_l
\end{array}
\right)=\left(\begin{array}{c}
0\\0
\end{array}
\right)
\end{eqnarray}
for $l\ge0$, and
\begin{eqnarray}\label{eq-118}
\left(\begin{array}{cc}
K_{l+n}(\tilde r_0/\lambda) & I_{l+n}(\tilde r_0/\lambda)\\
-K_{l+n+1}(\tilde R/\lambda) & I_{l+n+1}(\tilde R/\lambda)
\end{array}
\right)
\left(\begin{array}{c}
A_l\\ B_l
\end{array}
\right)=\left(\begin{array}{c}
0\\0
\end{array}
\right)
\end{eqnarray}
for $l\le-1$.  The solutions are non-vanishing if the determinant of
previous matrices is zero. Thus, we have the conditions (after
shifting indices and defining $z=\lambda^{-1}$)
\begin{eqnarray}\label{eq-120.}
K_{l+n+1}(\tilde r_0z)I_{l+n}(\tilde Rz)+K_{l+n}(\tilde Rz)I_{l+n+1}(\tilde r_0z)
\!\!\!&=&\!\!\!0,\,\,\,\,l\geq0,\\
K_{l+n-1}(\tilde r_0z)I_{l+n}(\tilde Rz)+K_{l+n}(\tilde Rz)I_{l+n-1}(\tilde r_0z)
\!\!\!&=&\!\!\!0,\,\,\,\,l\leq0.\label{eq-120x}
\end{eqnarray}
We now again write $n$ as $n=k+\nu$, with $k\in\mathbb Z$ and $|\nu|<1$.
Our conditions now become
\begin{eqnarray}\label{eq-120t3f}
K_{l+\nu+1}(\tilde r_0z)I_{l+\nu}(\tilde Rz)+K_{l+\nu}(\tilde Rz)I_{l+\nu+1}(\tilde r_0z)
\!\!\!&=&\!\!\!0,\,\,\,\,l\geq k,\\
K_{l-\nu+1}(\tilde r_0z)I_{-l+\nu}(\tilde Rz)+K_{l-\nu}(\tilde Rz)I_{-l+\nu-1}(\tilde r_0z)
\!\!\!&=&\!\!\!0,\,\,\,\,l\geq-k.
\end{eqnarray}
After eliminating the negative $l$-modes in favor of the positive ones we obtain
\begin{eqnarray}\label{eq-120t3a53}
K_{l+|\nu|+1}(\tilde r_0z)I_{l+|\nu|}(\tilde Rz)+K_{l+|\nu|}(\tilde Rz)I_{l+|\nu|+1}(\tilde r_0z)
\!\!\!&=&\!\!\!0,\,\,\,\,l\geq |k|
\\
K_{l+|\nu|-1}(\tilde r_0z)I_{l+|\nu|}(\tilde Rz)+K_{l+|\nu|}(\tilde Rz)I_{l+|\nu|-1}(\tilde r_0z)
\!\!\!&=&\!\!\!0,\,\,\,\,|k|\geq l \geq0
\\
K_{l-|\nu|+1}(\tilde r_0z)I_{l-|\nu|}(\tilde Rz)+K_{l-|\nu|}(\tilde Rz)I_{l-|\nu|+1}(\tilde r_0z)
\!\!\!&=&\!\!\!0,\,\,\,\,l\ge1
\end{eqnarray}
From now on we will assume $\nu>0$ and $k>0$. Let us define the
functions
\begin{eqnarray}\label{eq-120t3a5}
h_l^{(1)}(z)\!\!\!&=&\!\!\!\frac{\tilde r_0^{l+\nu+1}z}{\tilde R^{l+\nu}}
\Big[K_{l+\nu+1}(\tilde r_0z)I_{l+\nu}(\tilde Rz)+K_{l+\nu}(\tilde Rz)I_{l+\nu+1}(\tilde r_0z)\Big]
,\,\,\,\,l\geq k,
\\
h_l^{(2)}(z)\!\!\!&=&\!\!\!\frac{\tilde R^{l+\nu}z}{\tilde r_0^{l+\nu-1}}
\Big[K_{l+\nu-1}(\tilde r_0z)I_{l+\nu}(\tilde Rz)+K_{l+\nu}(\tilde Rz)I_{l+\nu-1}(\tilde r_0z)\Big]
,\,\,\,\,k\geq l \geq0,
\\
h_l^{(3)}(z)\!\!\!&=&\!\!\!\frac{\tilde r_0^{l-\nu+1}z}{\tilde R^{l-\nu}}
\Big[K_{l-\nu+1}(\tilde r_0z)I_{l-\nu}(\tilde Rz)+K_{l-\nu}(\tilde Rz)I_{l-\nu+1}(\tilde r_0z)\Big]
,\,\,\,\,l\geq1.
\end{eqnarray}
Using the well known properties of the Bessel functions~\cite{Grad},
we can check the conditions $h_l^{(i)}(0)=1$, $h_l'^{(i)}(0)=0$, and
$h_l^{(i)}(z)=h_l^{(i)}(-z)$. As similarly as performed in
\cite{ref-corn,siete,Merchan-Tellez-jabon-anillos,ferrero}, those
properties allow us to decompose our grand potential as a Weierstrass
product over its zeroes. The grand potential for our system is
\begin{equation}
\beta\Omega=-\sum_{i=1}^{3}\sum_l\ln h_l^{(i)}(-1),
\end{equation}
where $l$ lies in the intervals $[k,\infty)$, $[0,k]$, and $[1,\infty)$ for $i=1,2$, and $3$, respectively.

Therefore, we get
\begin{eqnarray}\label{eq-120}
\beta\Omega\!\!\!&=&\!\!\!
\beta\Omega_0+\beta\Omega(n,\tilde R)+\beta\Omega(k,\nu,\tilde r_0)
+\beta\Omega^{\,\nu\!}(\tilde R)
+O(e^{-2\tilde R}),
\end{eqnarray}
where \cite{ref-corn}
\begin{eqnarray}\label{eq-120a}
\beta\Omega_0\!\!\!&=&\!\!\!
-\pi R^2\,\frac{m^2}{2\pi}\Big[\ln\frac{_2}{^{ma}}-\gamma+\frac{_1}{^2}\Big]
+2\pi R\,m\Big[\frac{_1}{^4}-\frac{_1}{^{2\pi}}\Big]
+\frac{_1}{^6}\ln(mR)+O(1)
\end{eqnarray}
is the grand potential of the unperturbed plasma, and (remember that $k+\nu=n$)
\begin{eqnarray}\label{eq-120w7}
\beta\Omega(n,\tilde R)\!\!\!&=&\!\!\!
n\ln(mR)-\ln[\,\Gamma(n+1)\,]-n\ln2+O(1/\tilde R),
\\\label{eq-120w8}
\beta\Omega(k,\nu\,,\tilde r_0)\!\!\!&=&\!\!\!
-\!\sum_{l=k}^{\infty}\ln\!\Bigg[\left(\frac{\tilde r_0}{2}\right)^{l+\nu+1}
\!\!\!\!\!\frac{2\,K_{l+\nu+1}^{(0)}}{\Gamma(l+\nu+1)}\Bigg]
-\!\sum_{l=1}^{\infty}\ln\!\Bigg[\left(\frac{\tilde r_0}{2}\right)^{l-\nu+1}
\!\!\!\!\!\frac{2\,K_{l-\nu+1}^{(0)}}{\Gamma(l-\nu+1)}\Bigg]
\nonumber\\
\!\!\!&&\!\!\!-\!\sum_{l=0}^{k}\ln\!\Bigg[\frac{\tilde R^{2(l+\nu)}}{\tilde r_0^{l+\nu-1}}
\frac{\,K_{l+\nu-1}^{(0)}}{2^{l+\nu}\Gamma(l+\nu+1)}\Bigg],
\\
\beta\Omega^{\,\nu\!}(\tilde R)\!\!\!&=&\!\!\!
-\sum_{l=0}^{\infty}\ln\!\left[\frac{I_{l+\nu}^{(R)}}{I_{l}^{(R)}}
\frac{\Gamma(l+\nu+1)}{\Gamma(l+1)}
\!\left(\frac{2}{\tilde R}\right)^{\nu}\right]
-\sum_{l=1}^{\infty}\ln\!\left[\frac{I_{l-\nu}^{(R)}}{I_{l}^{(R)}}
\frac{\Gamma(l-\nu+1)}{\Gamma(l+1)}
\!\left(\frac{2}{\tilde R}\right)^{-\nu}\right]
\nonumber\\
\!\!\!&&\!\!\!
-\ln\left[\frac{I_{\nu}^{(R)}}{I_{0}^{(R)}}\right].
\end{eqnarray}
Last expressions cannot be evaluated exactly. For
$\tilde{r}_0\ll 1$ and $\tilde{R}\gg 1$ we can use the Euler Mc-Laurin
formula to transform discrete sums into integrals and
expand $K_l^{(0)}$ in powers of $\tilde r_0$ using the fact that
$\tilde r_0\ll1$. After some algebra, adding all the different
contributions so we can write the grand potential as
$\beta\Omega=\beta\Omega_0+\beta\Omega_1(k,\nu)$, and neglecting
$O(1)$ terms we find

\begin{eqnarray}\label{131f2}
\beta\Omega_1(0,0)\!\!\!&=&\!\!\!
\frac{_{\tilde r_0^2}}{^{2}}\ln\Big(\frac{_R}{^a}\Big)
+O(\tilde r_0^2\ln\tilde r_0),
\\
\beta\Omega_1(n,0)\!\!\!&=&\!\!\!
-\ln\Big[\ln\Big(\frac{_{2}}{^{\tilde r_0}}\Big)-\gamma\,\Big]+n(n-1)\ln\tilde r_0-n^2\ln(mR)
+\frac{_{\tilde r_0^2}}{^{2}}\ln\Big(\frac{_R}{^a}\Big)
+O(\tilde r_0^2\ln\tilde r_0),
\nonumber\\
\end{eqnarray}
and
\begin{eqnarray}\label{eq-168}
\beta\Omega_1(0,\nu)\!\!\!&=&\!\!\!-\nu^2\ln(mR)
+\frac{_{\tilde r_0^2}}{^2}\ln\Big(\frac{_R}{^a}\Big)
+O(\tilde r_0^{2(1-\nu)}),
\\
\beta\Omega_1(k,\nu)\!\!\!&=&\!\!\!
\Big[n(n-1)+\nu(1-\nu)\Big]\ln\tilde r_0-n^2\ln(mR)
+\frac{_{\tilde r_0^2}}{^{2}}\ln\Big(\frac{_R}{^a}\Big)
+O(\tilde r_0^{2(1-\nu)},\,\tilde r_0^{2\nu}).\,\,\,\,
\end{eqnarray}
The results found for the grand potential are consistent with what we
expect. First of all, the bulk pressure and the surface tension are
not modified by the presence of the impurity, because we are working
in the thermodynamic limit and one single impurity cannot alter
extensive quantities, or quantities proportional to the boundary
length of the system. Second, we expect to find terms proportional to
$\ln\tilde r_0$ that diverge when $r_0\rightarrow0$ and increase with
$n$ because of the mentioned particle collapse at $\Gamma=2$. However,
notice that such divergence does not appear when $n=0$ because such
condition only applies for charged particles; the only contribution in
the case $n=0$ is a hard core effect that depends on the cutoff $a$
and vanishes as $r_0\rightarrow0$, Eq. (\ref{131f2}), which is
expected. Notice that such hard core effect is present for all values of
$n$.

The term proportional to $\ln(mR)$ is a universal finite-size
correction term that is related to the topology of the considered
system. Generally, for a two-dimensional conformal field theory,
confined in a domain of typical length $R$, in the limit $R\to\infty$,
the free energy of the system exhibits a finite size correction 
given by $\beta\Delta F=-\frac{1}{6}c\chi\ln(R)$, where $c$ is the
conformal anomaly number (central charge) and $\chi$ the Euler
characteristic of the domain containing the system \cite{cardi}. As
explained in~\cite{mani, jancotelcoulcrit2}, Coulomb systems should
exhibit a similar correction, with $c=1$, due to the long range of the
electric potential correlations, however with a change of sign, due to
the fact that, in the partition function, one integrates over the
fluctuations of the density and not directly over the fluctuations of
the electric potential. Now, in the present system, we have shown that
the finite size correction to the grand potential is changed from
$\frac{1}{6}\ln(mR)$ to $\frac{1-6n^2}{6}\ln(mR)$. The presence of the
charged impurity modifies the central charge from $c=1$ to
$c=1-6n^2$. This is in agreement with what is expected from conformal
field theory, as a simple deformation of the minimal free boson
conformal field theory ($c=1$) is obtained by spreading out a charge
$\alpha_0/2=n$ at infinity to obtain a conformal field theory with
$c=1-24\alpha_0^2=1-6n^2$~\cite{dotsenko-conforme, ginsparg}. For the
Coulomb system studied here, the external charge is not spread at
infinity, but located at the origin. Nevertheless it has the same
effect of shifting the central charge from $c=1$ to $c=1-6n^2$.

One difference between the contributions of integer and non-integer
charges to the grand potential is in the terms associated with $\ln
\tilde r_0$. While the term $\Delta\Omega=n(n-1)\ln\tilde r_0$ is
common in both cases, there is an additional contribution that differs in
the integer and the non-integer cases. For integer charges, this
contribution is
$\beta\Delta\Omega_{int}=-\ln\left[\ln\left(\frac{2}{\tilde
    r_0}\right)-\gamma\right]$, which clearly differs from the
contribution for non-integer charges, given by
$\beta\Delta\Omega_{non-int}=\nu(1-\nu)\ln\tilde r_0$.

\section{Conclusions}\label{conc}

We have studied a two-dimensional two-component plasma at $\Gamma=2$
with an electric impurity confined in a large disk of radius $R$.
Particularly, we found analytical expressions for the density and
correlation functions and the grand potential, which provide
information for the pressure and the superficial tension. When the
impurity is located at the origin of the confined plasma there is
rotational invariance and the equations can be solved analytically
using the method described in \cite{ref-corn}.  The counter-ions and
co-ions accumulate close to the origin and the boundary
respectively. 

The case in which the electric charge of the impurity is an integer
multiple of the charges in the plasma $\pm e$ and the case where the
impurity charge is not an integer multiple of $\pm e$ were analyzed
independently, to highlight some interesting differences due to the
discrete nature of the electric charges of ions of the plasma.
Although we could have expected quite different behaviors in the cases
where the charge takes an ``integer'' and a ``non-integer'' value, we
found that the effects on the charge redistribution are similar when
the radius of the impurity $r_0\neq 0$. But for a point-like impurity,
$r_0=0$, important differences can be observed between the ``integer''
and ``non-integer'' cases. When the impurity charge has an integer
value and $r_0\to0$ there is an integer number of counter-ions that
can cancel such effect when they collapse at the site where the
impurity is located. On the other hand, finite-size effects for finite
$r_0$ avoid the cancellation of the charge of the impurity and so the
charge is differently redistributed close to the impurity.  When
$r_0>0$ the charges redistribute in such way that there are
accumulated charges $-n$ and $n$ around the impurity and the boundary
respectively. In the case $r_0\to0$, and a non integer value of the
charge $n=k+\nu$ ($\nu\neq 0$), $k$ counter-ions collapse into the
impurity, and the charge accumulated around the impurity is only
$-\nu e$.  

This problem can be extended with the introduction of an additional
impurity. Unfortunately, the rotational symmetry in this situation is
broken and it might not be possible to find analytical results. When
two impurities are included, we expect to obtain the effect of each
individual particle (ignoring the effects of the other) plus an
additional term that is related to the interaction between the two
particles.

The results for the grand potential are as expected. Since the size of
the impurity is negligible compared to the the size of the disk, we do
not expect the pressure and the superficial tension to be
modified. Nonetheless, the central charge of the system is modified
because of the presence of the impurity. In this case, we found that
the topological term takes the form
$\left(\frac{1}{6}-n^2\right)\ln(mR)$, with $n$ the charge of the
impurity, indicating a change in the central charge from $c=1$ to
$c=1-6n^2$.

Partial financial support from Fondo de Investigaciones, Facultad de
Ciencias, Universidad de los Andes (project 2014-2 ``Impurezas
cargadas en plasmas y electrolitos''), and
ECOS-Nord/COLCIENCIAS-MEN-ICETEX is acknowledged.


\end{document}